\documentclass[aip,jcp,reprint,groupedaddress,showpacs]{revtex4-1}
\usepackage[T1]{fontenc}
\usepackage[utf8]{inputenc}
\usepackage{amsmath}
\usepackage{graphicx}
\usepackage{amssymb}
\usepackage{amsmath}
\usepackage{pstricks}
\usepackage{textcomp}
\usepackage{lineno}
\usepackage{mathrsfs}

\newcommand{\bvec}[1]{{\bf\string#1 }}
\newcommand{\upd}{\mathrm{d}}

\usepackage[normalem]{ulem}

\newcommand{\fu}[2]{\ensuremath{\left\{\begin{array}{l}#1\\#2\end{array}\right.}}

\newcommand{\vrx}[3]{\ensuremath{\left(\begin{array}{c}#1\\#2\\#3\end{array}\right)}}

\newcommand{\mat}[9]{\ensuremath{\left(\begin{array}{c}#1\\#4\\#7\end{array}\begin{array}{c}#2\\#5\\#8\end{array}\begin{array}{c}#3\\#6\\#9\end{array}\right)}}

\newcommand{\fuxyC}[4]{\ensuremath{\left.\begin{array}{l}#1\\#2\\#3\\#4\end{array}\right.}}

\begin{document}

\title{Surface tension of isotropic-nematic interfaces: Fundamental Measure Theory for hard spherocylinders}

\author{Ren\'e Wittmann}
\author{Klaus~Mecke}
\email{Klaus.Mecke@physik.uni-erlangen.de}
\affiliation{\mbox{$^1$ Institut f\"ur Theoretische Physik, Universit\"at Erlangen-N\"urnberg, Staudtstr.~7, D-91058 Erlangen, Germany}}

\date{\today}

\begin{abstract}
A fluid constituted of hard spherocylinders is studied using a density functional theory for non-spherical hard particles, which can be written as a function of weighted densities. This is based on an extended deconvolution of the Mayer $f$-function for arbitrarily shaped convex hard bodies in {\it tensorial} weight functions, which depend each only on the shape and orientation of a single particle. In the course of an examination of the isotropic-nematic interface at coexistence the functional is applied to anisotropic and inhomogeneous problems for the first time. We find good qualitative agreement with other theoretical predictions and also with Monte-Carlo simulations. 

\end{abstract}


\maketitle

\section{Introduction}
\label{sec_introduction}

Fluids of non-spherical particles can spontaneously align at sufficiently high densities or low temperatures \cite{degennes93}. These liquid crystals are used nowadays for many technological devices, since the direction of their preferred orientation can be tuned easily by external fields. In his seminal work \cite{On49} Onsager showed that a system composed alone of hard elongated particles can undergo a first-order phase transition from an isotropic to a nematic phase. The stability of the orientational order is solely due to entropic reasons as the particles only interact via hard-core repulsion. It is related to packing effects at increasing densities. Although in real systems attractive forces between the particles play an important role, in particular for temperature dependence of physical quantities, the hard-core repulsion alone can explain the main features of liquid crystals.

The Onsager theory for rods of infinite length has been successfully applied
to the coexistence of isotropic and nematic bulk fluids \cite{Ons1,Ons2,Ons3} and inhomogeneous systems \cite{chen92,SR}. 
However, this approach fails in
the description of hard rods with a finite length, especially for low
aspect ratios \cite{Stra73}. The breakthrough in the theoretical
description of inhomogeneous fluids came in 1979 when classical
Density functional theory (DFT) \cite{DFT} emerged. It enabled more
sophisticated calculations beyond the Onsager second virial approximation
and hence a better description of shorter rods. Parsons and Lee
\cite{Parsons,SDL} incorporated the virial series of hard spheres and
introduced a decoupling between translational and orientational degrees of
freedom. The successful weighted density approach has been adapted by
Poniewierski and Holyst \cite{PH,PH1,PH2} and Somoza and Tarazona
\cite{ST,ST2,ST3}. The latter density functional has been applied to more complex systems
with an improved computational evaluation by successors
\cite{VMS,velasco02}. It appears to be very accurate for
inhomogeneous problems as it is based on Tarazona's original functional for hard
spheres \cite{TarazonaE,Tarazona}. The most elaborate approach for hard spheres has
been made in Rosenfeld's Fundamental measure theory (FMT) \cite{Rf89}
which includes a whole set of weighted densities.

Understanding the properties of the isotropic-nematic interface
remained an interesting problem despite the simplicity of the hard body
model. The reasons are at least threefold: Experiments indicate values
smaller than $\gamma_{\text{IN}}\approx 10^{-3}$mN/m for the interfacial tension
\cite{chen2002} which would be even lower for particles without 
attractions. Thus accurate computer simulation becomes difficult. 
Recent simulations have been done for spherocylinders \cite{vink05,wolfsheimer06,schmid07,Allen0} or ellipsoids
\cite{Allen1,Allen2}. An early mean-field theory
\cite{kimura85,kimura86,kimura93} for fluids of hard rods with attractive
as well as repulsive interactions captures the qualitative behavior but
fails in quantitative predictions - in particular for purely repulsive
hard-particle fluids. Few density functionals have been applied to
this problem beyond the Onsager approximation. The introduction of an artificially
sharp interface induces spurious minima in the interfacial tension as a
function of the tilt angle $\varTheta$ between the director and the interface normal
\cite{holyst88}. The so far most advanced DFT study has been carried out
with a free minimization of the Somoza-Tarazona functional
\cite{velasco02}. Its main observations are that the interfacial tension
is a monotonically decreasing function of $\varTheta$ and that there is a
shift between the density profile and the profile of the nematic order
parameter which are shaped like hyperbolic tangents. Although these
qualitative results coincide with computer simulations
\cite{vink05,wolfsheimer06,schmid07}, the quantitative significance of the
calculated values remains unsure.

The nematic surface at a hard wall as well as the interface between the
isotropic and nematic phase are notorious difficult problems, mainly
related to anisotropic steric excluded-volume interactions. The
decomposition of this hard-core interaction by applying the Gauss-Bonnet
theorem is one of the the main features employed in this paper. A free energy density functional for inhomogeneous hard-body fluids was derived in Refs.~\onlinecite{hansengoos09,hansengoos10} on this
foundation. It can describe a
stable nematic phase and an isotropic-nematic transition for the
hard-spherocylinder fluid in contrast to previous functionals of its
kind. The new functional also improves in the description of inhomogeneous
isotropic fluids when comparing with data from Monte-Carlo simulations for
hard spherocylinders in contact with a planar hard wall. In this paper, we
continue this study by the following steps:

First we recapitulate in Sec.~\ref{sec_DFT} the \emph{extended deconvolution Fundamental measure theory} (edFMT)\cite{hansengoos09,hansengoos10} for inhomogeneous hard-body fluids, which reduces to Rosenfeld's FMT \cite{Rf89} when applied to hard spheres. 
In Sec.~\ref{sec_interface} we apply this functional to homogeneous fluids of hard spherocylinders with length $L$ and diameter $D$ and show that it captures the isotropic-nematic transition. An explicit expression for the surface tension is derived within a Landau-de Gennes theory for hard rod interfaces. Section \ref{sec-surfacetension} provides a study of the isotropic-nematic interface where we calculate the interfacial tension within DFT and conclude with a discussion of our results in comparison with computer simulations \cite{vink05}.

\section{Tensorial fundamental measure theory} 
\label{sec_DFT} 

The FMT functional as introduced by Rosenfeld \cite{Rf89}, together with improvements concerning the underlying equation of state \cite{RoEvLaKa02,YuWu02,HaRo06} as well as highly confined geometries \cite{RfSchLoeTar96,RfSchLoeTar97,TarRf97,Tar99}, is the most successful DFT for polydisperse mixtures of hard spheres. Its simplicity comes from exclusively including geometrical measures of hard spheres without empirical inputs. Despite the success of this functional an adequate generalization to anisotropic hard bodies has been missing for a long time. The proposition of Rosenfeld \cite{Rf94,Rf95} fails to describe nematic ordering and the DFT by Cinacchi and Schmid \cite{CiSchm02} is not constructed with one-center convolutions. Other functionals were not derived for arbitrarily shaped bodies \cite{esztermann06,martinezraton08}. 
Finally the problem has been resolved in 2009 by an \textit{extended deconvolution} of the Mayer $f$-function which gives rise to an appropriate functional for nematic order \cite{hansengoos09}. In the following we give an introduction to the essentials of this edFMT closely following the work of Rosenfeld \cite{Rf89}. 
Within the framework of DFT \cite{DFT} the grand potential functional
\begin{eqnarray}
\label{functional} 
  \Omega[\{\rho_i\}] & = & \mathcal{F}_{\mathrm{id}}[\{\rho_i\}] +   \mathcal{F}_{\mathrm{ex}}[\{\rho_i\}] \\ 
 &  &+ \sum_{i=1}^{\kappa}\int\!\upd\bvec{r} \int \mathrm{d}\varpi \  \rho_i(\bvec{r,\varpi}) (V_i^{\mathrm{ext}}(\bvec{r,\varpi})-\mu_i)  \nonumber 
\end{eqnarray}
 of a $\kappa$-component fluid of hard bodies $\mathcal{B}_i$ with orientation $\varpi$ and center $\bvec{r}$ can be separated into the free energy
\begin{equation}
  \beta \mathcal{F}_{\mathrm{id}}
  = \sum_{i=1}^{\kappa}\int\!\upd\bvec{r} \int \mathrm{d}\varpi \   \rho_i(\bvec{r},\varpi) (\ln(\rho_i(\bvec{r},\varpi)\Lambda_i^3)-1 )  
\end{equation}
 of an ideal gas where $\beta^{-1}=k_{\mathrm{B}}T$ is the inverse temperature
and the excess (over ideal gas) free energy $\mathcal{F}_{\mathrm{ex}}$ which contains the explicit interactions between the particles. Both are functionals of the orientational-dependent average particle number densities $\rho_i(\bvec{r},\varpi)$ of species $i=1\ldots\kappa$ with chemical potential $\mu_i$ and thermal wavelength $\Lambda_i$. The external potential $V_i^{\mathrm{ext}}(\bvec{r},\varpi)$ acts on each species. The equilibrium density profiles can be calculated from the Euler-Lagrange equations ${\delta  \Omega[\{\rho_i\}]}/{\delta \rho_i} \equiv 0$ for a given functional $\Omega[\{\rho_i\}]$. 
In the spirit of FMT we derive the extrapolated excess free energy $\mathcal{F}_{\mathrm{ex}}$ from the building blocks of an exact low-density expression.

\subsection{Deconvolution of the Mayer $f$-function}

Within the theory of diagrammatic expansions \cite{HaMcDo86} the lowest order term of the excess free energy reads
\begin{eqnarray}
\label{eq_ldlimit}
   \lim_{\rho_i\to 0}&& \beta \mathcal{F}_{\mathrm{ex}}
   =  \lim_{\rho_i\to 0}  \int\!\upd\bvec{r}\; \Phi_{\text{ex}}(\bvec{r}) \\\nonumber
  && =   - \frac{1}{2} \sum_{i,j=1}^{\kappa}\iint \upd\mathcal{R}_1\: \upd\mathcal{R}_2 \: \rho_i(\mathcal{R}_1) \:\rho_j(\mathcal{R}_2)\: f_{ij}(\mathcal{R}_1, \mathcal{R}_2)  \, ,
\end{eqnarray}
 with $\mathcal{R} = (\bvec{r},\varpi)$. The characteristic function
\begin{equation}
\label{eq_mayfnonsph}
 f_{ij}(\mathcal{R}_1, \mathcal{R}_1) = \left\{ \begin{array}{cl} 0 & \quad {\rm if } \;\; \mathcal{B}_i \cap \mathcal{B}_j = \emptyset \\
      -1 & \quad {\rm if } \;\; \mathcal{B}_i \cap \mathcal{B}_j \neq \emptyset   \end{array}\right.
\end{equation}
of the interaction between two convex hard bodies $\mathcal{B}_i$ and $\mathcal{B}_j$ is called the Mayer $f$-function. It only depends on the distance $\bvec{r}_1-\bvec{r}_2$ and the relative orientation of these bodies via their intersection $\mathcal{B}_i \cap \mathcal{B}_j$. The idea of FMT is to exclusively write the interaction given by Eq.~(\ref{eq_mayfnonsph}) in geometric expressions, specifically in terms of convolution products
\begin{equation}
\label{eq_defconv}
  \omega_i^{(\nu)} \otimes \omega_j^{(\mu)} = \int\!\upd\bvec{r}\; \omega_i^{(\nu)}(\bvec{r}-\bvec{r}_1,\varpi_1)\:\omega_j^{(\mu)}(\bvec{r}-\bvec{r}_2,\varpi_2) 
\end{equation}
of the weight functions $\omega_i^{(\nu)}(\bvec{r},\varpi) $ which characterize the shape of a single convex body $\mathcal{B}_i$ with arbitrary orientation $\varpi$. 
The general, orientation-dependent scalars and vectors
\begin{eqnarray}
\label{weightfunctions}
   \omega_i^{(3)}(\mathcal{R}) &=& \Theta\left(|\mathbf{R}_i(\hat{\mathcal{R}})|-|\mathbf{r}|\right) \, , \cr
   \omega_i^{(2)}(\mathcal{R}) &=& \frac{\delta(|\mathbf{R}_i(\hat{\mathcal{R}})|-|\mathbf{r}|)}{ \bvec{n}_i(\hat{\mathcal{R}})\hat{\bvec{r}}} \, , \cr  
   \omega_i^{(1)}(\mathcal{R}) &=& \frac{H_i(\hat{\mathcal{R}})}{4\pi}\,\omega_i^{(2)}(\mathcal{R}) \, , \cr
  \omega_i^{(0)}(\mathcal{R}) &=& \frac{K_i(\hat{\mathcal{R}})}{4\pi}\,\omega_i^{(2)}(\mathcal{R}) \, , \cr
   \overrightarrow{\omega}_i^{(2)}(\mathcal{R}) &=&  \mathbf{n}_i(\hat{\mathcal{R}}) \, \omega_i^{(2)}(\mathcal{R}) \, , \cr
   \overrightarrow{\omega}_i^{(1)}(\mathcal{R}) &=& \frac{H_i(\hat{\mathcal{R}})}{4\pi} \overrightarrow{\omega}_i^{(2)}(\mathcal{R}) 
   \end{eqnarray}
   which are also present in the hard sphere functional \cite{Rf89} contain an additional factor $(\bvec{n}_i(\hat{\mathcal{R}})\hat{\bvec{r}})^{-1}$ which accounts for different parametrizations \cite{hansengoos09} and $\hat{\mathcal{R}}=(\hat{\bvec{r}},\varpi)$. The additional tensorial weight functions
      \begin{eqnarray}
\label{weightfunctions2}
  \overleftrightarrow{\omega}_i^{(2)}(\mathcal{R})  &=&  \bvec{n}_i(\hat{\mathcal{R}})   \bvec{n}_i(\hat{\mathcal{R}}) ^{\mathrm{T}}  \,  \frac{\delta(|\mathbf{R}_i(\hat{\mathcal{R}})|-|\mathbf{r}|)}{ \bvec{n}_i(\hat{\mathcal{R}})\hat{\bvec{r}}} \, ,\cr
 \overleftrightarrow{\omega}_i^{(1)}(\mathcal{R}) &=& \frac{{\Delta\kappa}_i(\hat{\mathcal{R}})}{4\pi} \left( \bvec{v}_i^I(\hat{\mathcal{R}}) \bvec{v}_i^I(\hat{\mathcal{R}})^{\mathrm{T}} - \bvec{v}_i^{I\!I}(\hat{\mathcal{R}}) \bvec{v}_i^{I\!I}(\hat{\mathcal{R}})^{\mathrm{T}} \right) \, \cr &&\times\frac{\delta(|\mathbf{R}_i(\hat{\mathcal{R}})|-|\mathbf{r}|)}{ \bvec{n}_i(\hat{\mathcal{R}})\hat{\bvec{r}}} \,  ,  
\end{eqnarray}
of rank 2 are constructed with the dyadic product $\bvec{a}\bvec{b}^{\mathrm{T}}$ of two identical vectors. 
A point on the surface $\partial\mathcal{B}_i$ of the body is given by $\bvec{R}_i(\hat{\mathcal{R}})$ and the radial unit vector is $\hat{\bvec{r}}=\bvec{r}/|\bvec{r}|$. The three mutual perpendicular unit vectors $\bvec{n}_i$, $\bvec{v}_{i}^{I}$ and $\bvec{v}_{i}^{I\!I}$ denote the outward normal to $\partial\mathcal{B}_i$ and the directions of the two local principal curvatures $\kappa_i^{I}$ and $\kappa_i^{I\!I}$ respectively. The surface is characterized by its mean $H_i = \frac{1}{2}(\kappa_i^{I}+\kappa_i^{I\!I})$, Gaussian $K_i=\kappa_i^{I}\kappa_i^{I\!I}$ and deviatoric curvature ${\Delta\kappa}_i=\frac{1}{2}(\kappa_i^{I}-\kappa_i^{I\!I})$. 
The multiplication in Eq.~(\ref{eq_defconv}) includes the matrix product followed by the trace for rank 2 tensors and the scalar product for vectors as $\omega_i^{(\nu)}(\mathcal{R})$ denotes a weight function of unspecified rank. 
The implementation of the orientational dependence is discussed in appendix \ref{appendixA} for a cylindrical symmetric body. 

As already proposed by Rosenfeld \cite{Rf94,Rf95} the Gauss-Bonnet theorem 
$\int K_i \, \mathrm{d} A + \int\kappa^{\mathrm{g}}_i \, \mathrm{d} s =  2\pi\chi(\partial\mathcal{B}_i\cap\mathcal{B}_j)$ is applied in Ref.~\onlinecite{hansengoos09} to obtain an approximate deconvolution of the Mayer $f$-function
\begin{eqnarray}
  \label{eq_decompfij} 
&&\left. \begin{matrix}  0 \cr 1 \end{matrix} \right\}  =  -f_{ij}(\bvec{r}=\bvec{r}_1-\bvec{r}_2,\varpi_1,\varpi_2) \cr 
  = &&
  \int\limits_{\partial\mathcal{B}_i\cap\mathcal{B}_j} \frac{K_i}{4\pi}\; \mathrm{d} A_i + \int\limits_{\mathcal{B}_i\cap\partial\mathcal{B}_j} \frac{K_j}{4\pi}\; \mathrm{d} A_j
+ \int\limits_{\partial\mathcal{B}_i\cap\partial\mathcal{B}_j} \frac{\kappa_i^{\mathrm{g}} + \kappa_j^{\mathrm{g}}}{4\pi} \; \mathrm{d} s \cr
   = &&\int\limits_{\partial\mathcal{B}_i\cap\mathcal{B}_j} \frac{K_i}{4\pi} \; \mathrm{d} A_i  + \int\limits_{\partial\mathcal{B}_i\cap\partial\mathcal{B}_j} \frac{H_i}{4\pi} (1-\bvec{n}_i\bvec{n}_j) \frac{\upd s}{|\bvec{n}_i\times\bvec{n}_j|} \cr 
&& - \int\limits_{\partial\mathcal{B}_i\cap\partial\mathcal{B}_j} \frac{{\Delta\kappa}_i}{4\pi} \, \frac{(\bvec{v}_{i}^{I}\bvec{n}_j)^2
  -(\bvec{v}_{i}^{I\!I}\bvec{n}_j)^2}{(1+\bvec{n}_i\bvec{n}_j )}\frac{\upd s}{|\bvec{n}_i \times \bvec{n}_j|} \cr && + \, (i \leftrightarrow j) \cr
  \approx && \ \omega_i^{(0)} \otimes \omega_j^{(3)} + \omega_i^{(1)} \otimes \omega_j^{(2)} - \overrightarrow{\omega}_i^{(1)} \otimes \overrightarrow{\omega}_j^{(2)} \cr
  &  & - \zeta \overleftrightarrow{\omega}_i^{(1)} \otimes \overleftrightarrow{\omega}_j^{(2)} + (i \leftrightarrow j)
\end{eqnarray}
for non-spherical particles which is exact for spheres as the deviatoric curvature ${\Delta\kappa}$ and hence $\overleftrightarrow{\omega}^{(1)}$ become zero. The shortcut $(i \leftrightarrow j)$ repeats all terms with indices $i$ and $j$ exchanged. The main achievement of the calculation presented in Ref.~\onlinecite{hansengoos10} is that the result 
\begin{equation}
\label{eq_appAfinal}
  \kappa_i^{\mathrm{g}}+ \kappa_j^{\mathrm{g}} = H_i  \frac{1-\bvec{n}_i
  \bvec{n}_j}{|\bvec{n}_i \times \bvec{n}_j|} - {\Delta\kappa}_i  \frac{(\bvec{v}_{i}^{I}\bvec{n}_j)^2
  -(\bvec{v}_{i}^{I\!I}\bvec{n}_j)^2}{(1+\bvec{n}_i\bvec{n}_j ) |\bvec{n}_i
  \times \bvec{n}_j|}  +  (i\!\leftrightarrow\! j) 
\end{equation} 
 for the geodesic curvature which is a geometric quantity depending on the shape and position of \textit{both particles}, can be rewritten in geometric terms of one particle. This result can be used to decompose the Mayer $f$-function completely. It completes Rosenfeld's approximate decomposition for non-spherical particles \cite{Rf94,Rf95}. However, the last term of Eq.~(\ref{eq_appAfinal}) can only be deconvoluted by an expansion of the denominator. 
 For practical reasons the approximation $(1+\bvec{n}_i\bvec{n}_j)^{-1}=1-\bvec{n}_i\bvec{n}_j+(\bvec{n}_i\bvec{n}_j)^2+\ldots\approx\zeta$ is made. Otherwise an infinite number of additional tensorial weight functions with increasing rank has to be considered in Eq.~(\ref{eq_decompfij}) for the exact deconvolution of the Mayer $f$-function.

\subsection{Excess free energy density}

A basic idea of FMT is that the low-density limit in Eq.~(\ref{eq_ldlimit}) can be rewritten in a simple form involving weighted densities
\begin{equation}
\label{eq_weighdenorient}
  n_{\nu}(\bvec{r}) = \sum_{i=1}^{\kappa} \int\!\upd\bvec{r}\:\!' \int \mathrm{d}\varpi \  \rho_i(\bvec{r}\:\!',\varpi)\: \omega_i^{(\nu)}(\bvec{r}-\bvec{r}\:\!',\varpi) \, ,
\end{equation}
which, in contrast to the densities  $\rho_i(\bvec{r}_i,\varpi_i)$, are non-local quantities and constitute the building blocks of the theory. Inserting  Eq.~(\ref{eq_decompfij}) into the low-density limit, Eq.~(\ref{eq_ldlimit}) leads to the excess free energy density 
\begin{eqnarray}
\label{eq_ldextdeconv}
  \Phi_{\text{ed}}\left(\{n_{\nu}(\bvec{r})\}\right) & = &   n_0(\bvec{r}) n_3(\bvec{r}) + n_1(\bvec{r}) n_2(\bvec{r}) -  \overrightarrow{n}_1(\bvec{r})
  \overrightarrow{n}_2(\bvec{r}) \nonumber\\
 & & - \; \zeta \mathrm{Tr}\left[\overleftrightarrow{n}_1(\bvec{r})  \overleftrightarrow{n}_2(\bvec{r})\right] \; +\; \mathcal{O}(\rho^3) \, . 
\end{eqnarray} 
 To describe the dense fluid, i.e.\ rods of lower aspect ratios beyond the Onsager approximation or inhomogeneous phases, the higher order terms in Eq.~(\ref{eq_ldextdeconv}) have to be determined. There is strong motivation to extrapolate this excess free energy density towards finite particle densities to an expression which still is a {\em function} of these eight weighted densities. As long as no equation of state is used as an input (see, e.g., Ref.~\onlinecite{HaRo06}), there is a straightforward way to do so. An exact relation from scaled particle theory \cite{SPT} gives rise to 
 \begin{eqnarray}
\label{eq_PhiED}
  \Phi_{\text{ed}} & = &  - n_0 \ln (1-n_3)  
  + \frac{n_1 n_2 - \overrightarrow{n}_1\overrightarrow{n}_2 - \zeta \mathrm{Tr}[\overleftrightarrow{n}_1  \overleftrightarrow{n}_2] }{1-n_3}\cr & &+ \frac{\phi_3(n_2,\overrightarrow{n}_2, \overleftrightarrow{n}_2 )}{(1-n_3)^2}\, ,
\end{eqnarray} 
where the arguments of the functions were omitted for convenience. The expression $\phi_3(n_2,\overrightarrow{n}_2, \overleftrightarrow{n}_2 )$ only depends on those three weighted densities due to dimensional considerations \cite{RemDim} and compatibility to Eq.~(\ref{eq_ldextdeconv}). For a hard sphere fluid with $\overleftrightarrow{n}_1\equiv0$ Eq.~(\ref{eq_PhiED}) results in the original Rosenfeld functional \cite{Rf89}. The best choice for the third term is not obvious when fluids of anisotropic hard bodies are considered. The original expression reads
\begin{equation}
\label{eq_phiRos}
\phi_3^{(\mathrm{RF})}(n_2,\overrightarrow{n}_2) = \frac{1}{24\pi}(n_2^3-3 n_2 \overrightarrow{n}_2 \overrightarrow{n}_2 )
\end{equation}
 and the term 
    \begin{figure} [t]
\centering
\includegraphics[height=0.28\textwidth] {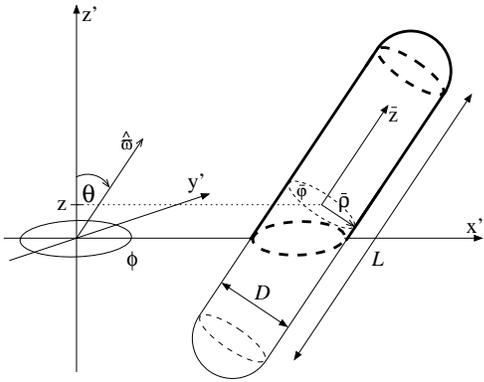}
\caption{Sketch of an oriented spherocylinder of length $L$ and diameter $D$ within a space-fixed coordinate system. A convex body $\mathcal{B}_i$ can be parametrized by a vector $\bvec{R}_i\left(\hat{\bar{\bvec{r}}}\right)$ which connects the center of mass 
of $\mathcal{B}_i$ with a point of the surface $\partial\mathcal{B}_i$. Here this is done in cylindrical coordinates $\bar{\bvec{r}}=(\bar{z}, \bar{\varrho},\bar{\varphi})$. The $\bar{z}$-axis is chosen to be parallel to the arbitrary orientation $\hat{\varpi}$ which is given by the two rotation angles $\theta$ and $\phi$. The thick lines on the surface indicate the parts of a spherocylinder centered at $z\:\!'=z$ which contribute to the thresholded weight functions $\omega_{\text{th}}^{(\nu)}(z,\varpi)$ defined in Eq.~(\ref{thres_wf}).}
  \label{fig_bodypar}
\end{figure}
 \begin{eqnarray}
\label{eq_phiTar}
  & \phi_3^{(\mathrm{T})}&(n_2,\overrightarrow{n}_2, \overleftrightarrow{n}_2 ) \\
&=& \!\frac{3}{16\pi}\left( \overrightarrow{n}_2^{\mathrm{T}} \overleftrightarrow{n}_2
  \overrightarrow{n}_2 - n_2 \overrightarrow{n}_2  \overrightarrow{n}_2 - \mathrm{Tr}[\overleftrightarrow{n}_2^3] + n_2 \mathrm{Tr}[\overleftrightarrow{n}_2^2] \right)  \nonumber 
\end{eqnarray}
 has been introduced by Tarazona \cite{Tar99} as the final result of dimensional crossover \cite{RfSchLoeTar96,RfSchLoeTar97,TarRf97,Tar99} to describe inhomogeneous hard sphere systems. This substitution dramatically improves the description of the crystal, which is never stable for the original Rosenfeld functional~\cite{Rf89}. Relatedly, it predicts a negative divergence of the free energy for a single cavity in the zero-dimensional limit~\cite{RfSchLoeTar97}.
 % This substitution dramatically improves the description of the crystal, which is overstabilized by the original Rosenfeld functional \cite{TarRf97} as the diverging free energy at close packing has the wrong sign. 
 The fluid phase of hard spheres is invariant as $\overleftrightarrow{n}_2=\frac{1}{3} n_2\mathbb{I}$, where $\mathbb{I}$ is
the unit matrix. Note that Eq.~(\ref{eq_phiTar}) has been introduced without the weighted density $\overleftrightarrow{n}_2$ appearing within the derivation of the original functional. 
 Now, within the generalized expression of edFMT, this weighted density is contained intrinsically. This motivated the consistent choice of taking Eq.~(\ref{eq_phiTar}) instead of Eq.~(\ref{eq_phiRos}) for the final edFMT functional \cite{hansengoos09}. The weighted densities for a one component homogeneous bulk fluid of spherocylinders (see Fig.~\ref{fig_bodypar}) with length $L$, diameter $D$ and volume $v$ read
\begin{eqnarray}
 n_3&=&\rho \left( \frac{\pi}{4}LD^2+\frac{\pi}{6}D^3 \right)  = \eta\, , \cr
 n_2 &=& \rho\!\:(\pi LD+\pi D^2)\, ,\cr
n_1 &=& \rho\left(\frac{L}{4}+\frac{D}{2}\right)\, ,\ \ \ n_0= \rho \, , \cr
(\overleftrightarrow{n}_2)_{11}&=&(\overleftrightarrow{n}_2)_{22}=\rho\left(\frac{\pi}{6}LD(2+S)+\frac{\pi}{3}D^2\right)\, ,\cr
 (\overleftrightarrow{n}_2)_{33} & =& \rho\left(\frac{\pi}{3}LD(1-S)+\frac{\pi}{3}D^2\right)\, ,\cr
\left(\overleftrightarrow{n}_1\right)_{11}&=&(\overleftrightarrow{n}_1)_{22}=-\frac{1}{2}(\overleftrightarrow{n}_1)_{33}=\rho\frac{L}{8}S\, , 
\label{eq_gewdich}
  \end{eqnarray}
with $\eta=\rho v$ the packing fraction and $S$ the nematic order parameter \cite{hansengoos10}. All important physical quantities calculated in Secs.~\ref{sec_interface} and \ref{sec-surfacetension} only depend on $\eta$ and the aspect ratio $l=L/D$. We use the functional based on Eq.~(\ref{eq_phiTar}) with $\zeta=1.6$ in our calculations if not denoted~otherwise.

\section{Isotropic-nematic interface} 
\label{sec_interface} 

We now turn to a study of bulk properties in the context of their influence on the isotropic-nematic interface. Sections \ref{sec_IN} and \ref{sec_isonemtrans} discuss the isotropic equation of state (EOS) and review the isotropic-nematic phase coexistence respectively. Section \ref{sec_LdG} introduces a Landau-de Gennes-theory for the isotropic-nematic interface.

\subsection{Homogeneous and isotropic fluid} \label{sec_IN} 
The isotropic phase, appears to be very well described. The exact second virial coefficient
\begin{equation}
\label{eq_B2}
  B_2\: \rho^2 = n_0 n_3 + n_1 n_2 - \overrightarrow{n}_1  \overrightarrow{n}_2 - \zeta \mathrm{Tr}[\overleftrightarrow{n}_1  \overleftrightarrow{n}_2] 
\end{equation}
for the homogeneous and isotropic bulk fluid defined by the relation $\Phi = B_2 \rho^2 + \mathcal{O}(\rho^3)$ is the same as for the Rosenfeld functional. This can be seen from the weighted densities in Eq.~(\ref{eq_gewdich}) as $\overleftrightarrow{n}_1=0$ for $S=0$. The isotropic EOS
\begin{equation}  
\beta p = -\Phi_{\text{ed}}+\sum_{i=0}^3\frac{\partial\Phi_{\text{ed}}}{\partial n_i}n_i+n_0  
\label{eq_PYwd}
\end{equation}
which results from Eq.~(\ref{eq_PhiED}) reads 
\begin{equation}  
\beta p=\frac{n_0}{1-n_3}+\frac{n_1n_2}{(1-n_3)^2}+\frac{n_2^3}{12\pi(1-n_3)^3}\, .
\label{eq_PYwd2}
\end{equation}
\begin{figure} [t]
\centering
\includegraphics[height=0.28\textwidth] {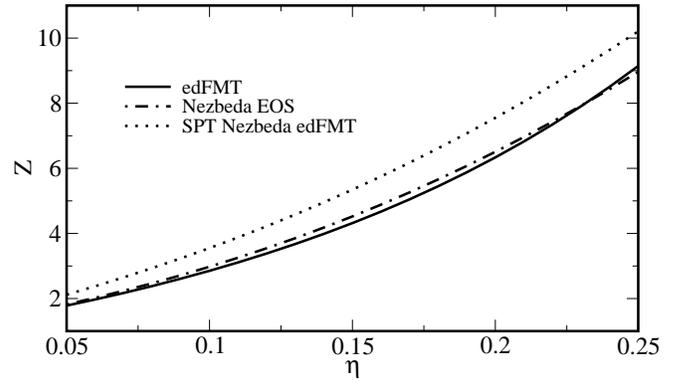}
\caption{Isotropic equation of state $Z=\beta p /\rho$ for hard spherocylinders with aspect ratio $l=L/D=10$ from the present functional (solid line) and Nezbeda (dot-dashed line) \cite{NEZ}. The dotted lines show the result $\beta p=\frac{\partial \Phi_{\text{ed}}}{\partial n_3}$ from scaled particle theory \cite{SPT} for an edFMT constructed with an imposed Nezbeda EOS according to Eq.~(\ref{eq_PYwd}).}
  \label{fig_EOS}
\end{figure}
Note that all vanishing terms of tensors and vectors are omitted. This result is obtained with both choices Eq.~(\ref{eq_phiRos}) and Eq.~(\ref{eq_phiTar}) for the third term as for the hard sphere fluid. By construction of Rosenfeld \cite{Rf88}, Eq.~(\ref{eq_PYwd2}) obeys the scaled particle relation $\beta p=\frac{\partial \Phi_{\text{ed}}}{\partial n_3}$ \cite{SPT} and yields a representation of the Percus-Yevick \cite{PY} EOS for hard spheres when choosing $L=0$. 
There were other successful efforts \cite{RoEvLaKa02,YuWu02,HaRo06} to implement the more sophisticated Carnahan-Starling EOS \cite{CS}, in particular its generalizations \cite{MCSL,WBII0} for mixtures. These \textit{White-Bear} versions may also be used with the weighted densities for anisotropic bodies. Note that the EOS arising from the \textit{White-Bear mark II} version of edFMT \cite{HaRo06} does not differ significantly from the EOS defined by Eq.~(\ref{eq_PYwd2}). One advanced EOS for monodisperse hard spherocylinders is given by Nezbeda \cite{NEZ} and can be written in terms of the weighted densities from Eq.~(\ref{eq_gewdich}) as
\begin{eqnarray}  
\beta p_{\text{Nez}} & = & \frac{\left(1-2n_3-n_3^2\right)n_0+\left(1+\frac{1}{3}n_3+\frac{4}{3}n_3^2\right)n_1n_2}{(1-n_3)^3}  \nonumber\\
 &  &  +\frac{\frac{1}{9}-\frac{5}{9}n_3}{(1-n_3)^3}\frac{n_1^2n_2^2}{n_0}\, .
\label{eq_nEOSwd}
\end{eqnarray}
The comparison for $l=10$ made in Fig.~\ref{fig_EOS} shows indeed some deviations between Eqs.~(\ref{eq_PYwd2}) and~(\ref{eq_nEOSwd}). We implemented the Nezbeda EOS i.e.\ terms proportional to $\frac{n_1^2n_2^2}{n_0}$ by substituting Eq.~(\ref{eq_nEOSwd}) into Eq.~(\ref{eq_PYwd}) and solving the differential equation in the spirit of Ref.~\onlinecite{HaRo06}. 
However, an improved functional was not obtained and the scaled particle differential equation could not be generally fulfilled which can be seen in Fig.~\ref{fig_EOS}. Attempts based on monodisperse spherocylinders led to  complex functionals restricted by further approximations. 
We choose not to carry on with this approach since the simple isotropic EOS is relatively well described and argue that it is much more important to find a good representation of the nematic EOS. This can not be achieved within an extrapolation to the functional in Eq.~(\ref{eq_PhiED}) based only on the scalar weighted densities. For the description of nematic order the tensorial weighted densities in Eq.~(\ref{eq_gewdich}) are vital. It is assumed that differences due to other expressions for $\phi(n_2,\overrightarrow{n}_2, \overleftrightarrow{n}_2 )$ in Eq.~(\ref{eq_PhiED}) are negligible as long as highly confined fluids are not considered \cite{hansengoos10}. In future work we need to clarify if that also true for other phases or different hard body fluids.
\begin{table}[t]
\centering
\begin{tabular}{clcccccc}
\hline \hline
$l$ & work & $\eta_{\text{I}}$ & $\eta_{\text{N}}$ & $S_{\text{N}}$ & $\gamma_{\text{IN}}^*$ & $\delta/L$ & $|\Delta z|/L$ \\ 
\hline 
5 & edFMT &  0.396 &  0.400 & 0.407 & 0.0159 &  1.29 & 0.84  \\
5 & DFT \cite{VMS,velasco02} &  0.400 &  0.417 &  & 0.0634 &  &  \\
5 & MC \cite{BoFr96} &  0.398 &  0.398 &  &  &  &  \\
10 & edFMT &  0.232 &  0.239 & 0.477 & 0.0263 &  1.02 & 0.56 \\
10 & DFT \cite{VMS,velasco02} &  0.251 &  0.276 &  & 0.0877 &  &  \\
15 & edFMT &  0.163 &  0.171 & 0.510 & 0.0329 &  0.94 & 0.49 \\
15 & MC \cite{vink05,wolfsheimer06,schmid07} &  0.173 &  0.198 & 0.7 & 0.10 & 0.71 & 0.37 \\
20 & edFMT &  0.126 &  0.134 & 0.531 & 0.0375&  0.90 & 0.46 \\
20 & DFT \cite{VMS,velasco02} &  0.143 &  0.164 &  & 0.114 &  &  \\
20 & MC \cite{BoFr96} &  0.139 &  0.171 & 0.808 &  &  &  \\
$\infty$ & edFMT  &  $2.700/l$ &  $3.151/l$ & 0.624  & 0.0641 &  0.76 & 0.37 \\
$\infty$ & $\zeta=5/4$ &  $3.504/l$  & $3.872/l$   & 0.574  & 0.0637  &  0.81 & 0.40 \\
$\infty$ & ON \cite{SR,SRPhD}
 &  $3.287/l$ &  $4.184/l$ & 0.792 & 0.156 & $0.66$ & 0.45\\\hline\hline
 \end{tabular}
  \caption{Results for the isotropic-nematic coexisting densities $\eta_{\text{I}/\text{N}}$ and corresponding nematic order parameter $S_{\text{N}}$ for different aspect ratios $l=L/D$ of hard spherocylinders. Also shown is the interfacial tension $\gamma_{\text{IN}}^*=\beta\gamma_{\text{IN}} (L+D)D$, the width $\delta$ of the interface and the distance $\Delta z$ between the inflection points of the density and order parameter profile. The edFMT results are calculated with $\zeta = 1.6$ and also with $\zeta = 5/4$ in the Onsager limit. We use Eq.~(\ref{eq_dens_ht2}) for the  density profile,  which minimizes the interfacial tension at tilt angle $\varTheta=0.5\pi$ with the exception of $l=\infty$ (see Fig.~\ref{fig_ITHT10}(c)). In the text we give the result for the absolute minimum. 
  A comparison is made to Monte-Carlo simulations for the phase transition \cite{BoFr96} and the interface \cite{vink05,wolfsheimer06,schmid07}. Theoretical results are from the Somoza-Tarazona DFT \cite{VMS,velasco02} and the exact DFT in the Onsager limit \cite{SR,SRPhD}. The interface width $\delta=\xi_{\text{I}}+\xi_{\text{N}}$ from Ref.~\onlinecite{SRPhD} is calculated from the given correlation lengths. }
 \label{tab_IN}
 \end{table}

\subsection{Isotropic-nematic phase-transition}
\label{sec_isonemtrans}

Nematic order occurs when entropy can be gained by orientational alignment. 
At sufficiently high density  the hard-core excluded-volume term $\mathcal{F}_{\mathrm{ex}}$ compensates the increasing free energy $\mathcal{F}_{\mathrm{id}}$ of the ideal gas in Eq.~(\ref{functional}). 
The tensorial weighted densities in Eq.~(\ref{eq_gewdich}) depend on the nematic order parameter  
$S=\int_0^1\mathrm{d}\cos\theta\; \left(\frac{3}{2}\cos^2\theta-\frac{1}{2}\right) \:g(\cos\theta)$
which is defined as the average second Legendre Polynomial with respect to the orientational distribution $g(\cos\theta)$. For symmetry reasons the density $\rho\!\:(\varpi) =\rho \:g(\cos\theta)$ is a function of the azimuthal angle $\theta$ only. A straightforward calculation shows that the orientational distribution function reads \cite{hansengoos10}
\begin{equation}
g(\alpha,\cos\theta)=\frac{\alpha}{\mathcal{D}(\alpha)}  \exp\left(-\alpha^2\left(1- \cos^2\theta\right) \right) 
\label{eq_ordis}
\end{equation}
with $\alpha^2 = -\,\frac{3}{2\rho}\,\frac{\partial\Phi}{\partial S}$ and Dawson's integral $\mathcal{D}(\alpha)$. The intrinsic order parameter $\alpha$ can be determined self-consistently and yields $S=\frac{2\alpha^2}{15}+O(\alpha^4)$. Minimal solutions for $\alpha>0$ correspond to a nematic phase while the isotropic phase is given by $\alpha=0$ and thus $g(\cos\theta)\equiv1$.
In Table \ref{tab_IN} we summarize the values of the isotropic $\eta_{\text{I}}$ and nematic $\eta_{\text{N}}$ coexisting densities together with corresponding nematic order parameter $S_{\text{N}}$ for important aspect ratios $l$. The use of $\zeta = 1.6$ is observed \cite{hansengoos09} to be the best fit to the simulation data by Bolhuis and Frenkel \cite{BoFr96}. The difference $\delta\eta_{\text{IN}}=\eta_{\text{N}}-\eta_{\text{I}}$ between coexisting densities is generally underestimated. Unfortunately, the edFMT results $c_{\text{I}}=2.70$ and $c_{\text{N}}=3.15$ for the concentration $4c=\rho\:\pi L^2D$ in the Onsager limit should perfectly agree with the values $c_{\text{I}}=3.29$ and $c_{\text{N}}=4.19$ from Ref.~\onlinecite{Ons3}. Notice that the values for $\zeta=5/4$ shown in Table \ref{tab_IN} are equal to the first order results of the iteration done in that work. 
This clearly points out the limitations of the $\zeta$ correction and suggests the use of higher order terms.
However, the present functional with $\zeta=1.6$ is the first generalization of FMT which allows a sensible description of the nematic phase and is still based on one-center convolutions. Thus the predictions of this functional for the isotropic-nematic interface are of great interest.

\subsection{Landau-de Gennes theory for hard rod interfaces}
\label{sec_LdG}
In a first step we consider the isotropic-nematic interface of hard rods in relation to their bulk phase behavior from a phenomenological point of view. In terms of the grand canonical potential $\Omega(T,V,\mu)=\beta^{-1}\int\mathrm{d}\bvec{r}\;\omega(\bvec{r},\mu)$ the bulk Landau-de Gennes expansion \cite{deGennes} can be written
\begin{eqnarray}  
\omega_{\text{b}}\left(\mu,\left[\overleftrightarrow{Q}\right]\right)=&&\;\omega_0 +\frac{3}{4}A(\mu)\,Q_{\alpha\beta}Q_{\beta\alpha}+\frac{3}{2}B\,Q_{\alpha\beta}Q_{\beta\gamma}Q_{\gamma\alpha}\cr &&+\frac{9}{8}C\,(Q_{\alpha\beta}Q_{\beta\alpha})^2+\mathcal{O}\left(Q^5\right)
\label{LdGallg}
\end{eqnarray}
where the expression 
\begin{equation}  
A(\mu)=-a\:\beta(\mu-\mu^*)
\label{Amu}
\end{equation}
depends linearly on the chemical potential $\mu$.  
Within this expansion the isotropic phase becomes unstable for $\mu=\mu^*$. The explicit expression 
\begin{equation}  
Q_{\alpha\beta}(\bvec{r})=Q(\bvec{r})
\left(\hat{n}_\alpha(\bvec{r})\hat{n}_\beta(\bvec{r})-\frac{1}{3}\delta_{\alpha\beta}\right)
\label{QtensDIR}
\end{equation}
for the order parameter tensor includes the director field $\hat{n}(\bvec{r})=(\sin\varTheta\cos\varPhi,\sin\varTheta\sin\varPhi,\cos\varTheta)^T$ which is parallel to the $z$-axis for $\varTheta=0$. Substitution into Eq.~(\ref{LdGallg}) yields
\begin{equation}  
\omega_{\text{b}}(\mu,[Q])-\omega_0=\frac{1}{2}A(\mu)\,Q^2+\frac{1}{3}B\,Q^3+\frac{1}{4}C\,Q^4+\mathcal{O}\left(Q^5\right)
\label{LdGscal}
\end{equation}
with the scalar order parameter $Q(\bvec{r})$. The conditions for the isotropic-nematic bulk phase coexistence
\begin{equation}  
\omega_{\text{b}}(\mu_{\text{c}},Q)-\omega_0=0\ \ \ \mbox{and} \ \ \ \frac{\partial\omega_{\text{b}}(\mu_{\text{c}},Q)}{\partial Q}=0
\label{PTLdG}
\end{equation}
and the coexisting density difference
\begin{equation}  
\frac{\delta\eta}{v}=-\frac{\partial\omega_{\text{b}}(\mu,Q)}{\partial (\beta\mu)}=\frac{1}{2}aQ^2
\label{eq_drhoQ}
\end{equation}
evaluated for the DFT values $Q=S_{\text{N}}$, $\delta\eta=\delta\eta_{\text{IN}}$, $\mu_{\text{c}}$ and $\mu^*$ uniquely determine the parameters
\begin{eqnarray}  
a=2\:\frac{\delta\eta_{\text{IN}}}{vS_{\text{N}}^2} \, , \ B=6\:\frac{\delta\eta_{\text{IN}}\:\beta(\mu_{\text{c}}-\mu^*)}{vS_{\text{N}}^3} \, ,\ C=-\frac{2B}{3S_{\text{N}}} \, .\ \ \ \ \ \ && 
\label{eq_LdGparV}
\end{eqnarray}
The study of inhomogeneous systems requires an elastic term $f_{\text{d}}$. For a one-dimensional profile of the scalar order parameter $Q=Q(z)$ one obtains
\begin{equation}  
f_{\text{d}}[Q]=b_\varTheta^2\left(\frac{\mathrm{d}Q}{\mathrm{d}z}\right)^2\, , \ \ \ b_\varTheta^2=\beta\:\frac{6L_1+L_2\left(1+3\cos^2\varTheta\right)}{18}
\label{eq_Fd2n}
\end{equation}
from Eq.~(\ref{QtensDIR}). The Landau parameters $L_1$ and $L_2$ can be related to the Frank elastic coefficients \cite{Frank} of the nematic phase at coexistence. The low-order limits of the analytic edFMT expressions yield $L_\epsilon(\eta_{\text{N}}, S_{\text{N}},l)$\cite{PREPEC}. We find $L_2>0$ so that the lowest value of $b_\varTheta$ is obtained for $\varTheta=\pi/2$. The interfacial tension at $\mu=\mu_{\text{c}}$ can be obtained from a minimization of the functional
\begin{eqnarray}  
\beta\gamma_{\text{IN}}
=\int_{-\infty}^{\infty}\mathrm{d}z\left(\omega_{\text{b}}(\mu_{\text{c}},[Q])-\omega_0+ 
b_\varTheta^2\left(\frac{\mathrm{d}Q}{\mathrm{d}z}\right)^2\right)\, . \ \ \ \ \ 
\label{eq_gammaLDG}
\end{eqnarray}
The equilibrium order parameter profile 
\begin{equation}  
Q(z)=\frac{S_{\text{N}}}{2}\left(1+\tanh\left(\frac{z}{2\xi_\varTheta}\right)\right)
\label{eq_profileS}
\end{equation}
for each director orientation is the solution of the integrated Euler-Lagrange equation
\begin{equation}  
\omega_{\text{b}}(\mu_{\text{c}},[Q])-\omega_0=b_\varTheta^2\left(\frac{\mathrm{d}Q}{\mathrm{d}z}\right)^2
\label{eq_EL}
\end{equation}
for appropriate boundary conditions. The characteristic length scale is given by the correlation length \cite{deGennes} 
\begin{equation} 
\xi_\varTheta=b_\varTheta \sqrt{\frac{v}{\delta\eta_{\text{IN}}\:\beta(\mu^*-\mu_{\text{c}})}}S_{\text{N}} \, . 
\end{equation}  
Inserting Eq.~(\ref{eq_profileS}) into Eq.~(\ref{eq_drhoQ}) yields the density profile
\begin{equation}  
\rho\!\:(z)
=\rho_{\text{I}}+\frac{\delta\eta_{\text{IN}}}{4v} \left(1+\tanh\left(\frac{z}{2\xi_\varTheta}\right)\right)^2\, .
\label{eq_profileE}
\end{equation}
The director-dependent interfacial tension
\begin{equation}  
\beta\gamma_{\text{IN}}=b_\varTheta\frac{\sqrt{\delta\eta_{\text{IN}}\:\beta(\mu^*-\mu_{\text{c}})}}{3\:\sqrt{v}}S_{\text{N}}\, .
\label{gammaR}
\end{equation}
is calculated from Eq.~(\ref{eq_gammaLDG}) after the substitution with Eq.~(\ref{eq_EL}). It is directly proportional to the elastic prefactor $b_\varTheta$. This means that parallel alignment to the interface is favored. Substituting the low-order elastic coefficients from edFMT (see Ref.~\onlinecite{PREPEC}) into $b_\varTheta$ at $\varTheta=\pi/2$ leads to an expression
\begin{eqnarray} \beta\gamma_{\text{IN}}=\sqrt{\frac{\zeta\: l^2\:(10l^2+39)\:\delta\eta_{\text{IN}}\:\beta(\mu^*-\mu_{\text{c}})}{21\:(1-\eta_{\text{N}})\:(2+3l)^3\:\pi^2}}\frac{S_{\text{N}}\:\eta_{\text{N}}}{D^2}\ \ \ \ \ \ \ 
\label{ITscale}
\end{eqnarray}
which only depends on bulk properties at isotropic-nematic coexistence. All parameters can easily be obtained from the edFMT functional. The value for $\mu^*$ can be adapted to fit either the point of instability of isotropic or nematic phase or the intermediate maximum in Eq.~(\ref{LdGscal}) at coexistence. In Sec. \ref{sub_dc} we will compare the results to DFT values.

\section{DFT results for surface tensions} 
\label{sec-surfacetension} 

In this section we address the description of the shape and director dependence of the isotropic-nematic interfacial tension. Section \ref{sub_sk} introduces the problem within a sharp-kink approximation of the interfacial profile. A more advanced parametrization is given in Sec.~\ref{sub_ht} and Sec.~\ref{sub_dc} concludes with a discussion of the results and the possible necessity of a more sophisticated free numerical minimization. Appendix \ref{appendixA} gives additional insight into the calculation of the inhomogeneous weighted densities.

\subsection{Weighted densities at sharp interfaces} \label{sub_sk}
\begin{figure} [t]
\centering
\includegraphics[height=0.3\textwidth] {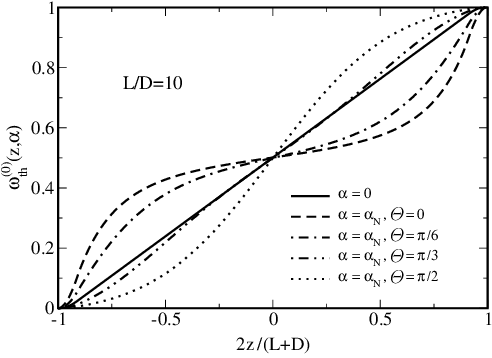}
\caption{Behavior of the average thresholded weight function $\omega_{\text{th}}^{(0)}(z,\alpha)$ as defined in Eq.~(\ref{thres_wfO}) for an aspect ratio of $l=10$. At the nematic coexisting order parameter $\alpha_{\text{N}}$ different values for the tilt angle $\varTheta$ are used. The isotropic  weight function with $\alpha=0$ does not depend on $\varTheta$. The function is constant for $|z|$ bigger than the half elongation $(L+D)/2$ of a spherocylinder.}
  \label{fig_wofz}
\end{figure}

\begin{figure} [bt]
\centering
\includegraphics[height=0.295\textwidth] {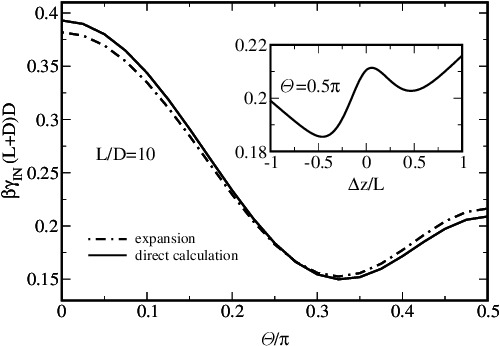}
\caption{Interfacial tension $\gamma_{\text{IN}}$ (solid line) as function of the tilt angle $\varTheta$ with an imposed sharp-kink profile, Eq.~(\ref{dens_sk}). The dot-dashed line is obtained from a quadratic expansion of the free energy. The aspect ratio of the hard spherocylinders is $l=10$. The minimum at $\varTheta\approx\pi/3$ does not appear in simulations but in a DFT calculation within the same approximation \cite{holyst88}. The inset shows the effect of a shift parameter $\Delta z$ between the density and order parameter profile. Negative values denote alignment at the isotropic side of the interface. }
  \label{fig_ITSK}
\end{figure}

A simple approximation describes the interface between the coexisting isotropic ($z<0$) and nematic ($z>0$) phase by a sharp-kink profile
\begin{equation}
\rho\!\:(z,\varpi) =\rho_{\text{I}} - \Theta(z)(\rho_{\text{I}}-\rho_{\text{N}}(\varpi))
\label{dens_sk}
\end{equation} 
which jumps from the homogeneous isotropic $\rho_{\text{I}}$ to the nematic $\rho_{\text{N}}(\varpi)$ coexisting density at $z=0$. Let us first define \textit{thresholded weight functions}
\begin{equation}  
\omega_{\text{th}}^{(\nu)}(z, \varpi)=\int\mathrm{d}\bvec{r}\:\!'\; \omega^{(\nu)}(\bvec{r}-\bvec{r}\:\!',\varpi)\: \Theta(z\:\!')
\label{thres_wf}
\end{equation}
of a single particle $\mathcal{B}$ centered at $z$ and its orientational average
\begin{equation}
\omega_{\text{th}}^{(\nu)}(z,\alpha) =\int\mathrm{d}\varpi \: \omega_{\text{th}}^{(\nu)}(z, \varpi) \:g_{\hat{n}}(\theta, \phi,\alpha)\, .
\label{thres_wfO}
  \end{equation}
  The generalized orientational distribution function 
  \begin{eqnarray}
g_{\hat{n}}(\alpha, \theta, \phi) & = & \frac{\alpha}{\mathcal{D}(\alpha)}\exp\left(-\alpha^2\left(1-\cos^2\vartheta\right)\right) \, , \\ 
 \cos\vartheta & := &  \sin\varTheta\sin\theta\cos\phi+\cos\varTheta\cos\theta
\nonumber   
\end{eqnarray}
  characterizes nematic order for an arbitrary nematic director $\hat{n}$ which includes an angle $\varTheta$ with the interface normal. As illustrated in Fig.~\ref{fig_bodypar} only the measures for $z\:\!'>0$ of a spherocylinder centered at $z$ contribute to the thresholded weight functions from Eq.~(\ref{thres_wf}). Thus for $z<-(L+D)/2$ all orientational averages in Eq.~(\ref{thres_wfO}) are zero while for $z>(L+D)/2$ one obtains the bulk weighted densities $n_{\nu}/\rho$ from Eq.~(\ref{eq_gewdich}). The behavior of $\omega_{\text{th}}^{(0)}(z,\alpha)$ shown in Fig.~\ref{fig_wofz} verifies the symmetry relation
    \begin{equation}
\omega_{\text{th}}^{(\nu)}\left(z>\frac{L+D}{2}, \alpha\right)=\omega_{\text{th}}^{(\nu)}(z, \alpha)+\sigma_\nu\:\omega_{\text{th}}^{(\nu)}(-z,\alpha)
  \end{equation} 
  with $\sigma_\nu=-1$ for the vectorial weights and $\sigma_\nu=1$ other\-wise. In terms of these thresholded weight functions the weighted densities $n_{i}(z)$ 
corresponding to the density profile from Eq.~(\ref{dens_sk}) read
\begin{equation}
n_{\nu}(z)=\rho_{\text{N}}\:\omega_{\text{th}}^{(\nu)}(z, \alpha_{\text{N}})+n_{\nu,\text{I}}-\rho_{\text{I}}\:\omega_{\text{th}}^{(\nu)}(z,0)   \, . 
\end{equation}
The surface tension between an isotropic and nematic phase with bulk pressure $p=p_{\text{I}}=p_{\text{N}}$ overall volume $V=V_{\text{I}}+V_{\text{N}}$ and interface area $A$ is defined by 
\begin{eqnarray}
\label{eq_gamma}
 & \gamma_{\text{IN}}&[\rho\!\:(z,\varpi)]  =   \frac{\Omega[\rho\!\:(z,\varpi)]+p_{\text{I}}V_{\text{I}}+p_{\text{N}}V_{\text{N}}}{A}  \\ 
&=&\int\!\!\mathrm{d}z\{\omega[\rho\!\:(z,\varpi)](z)-\omega[\rho_{\text{I}}]\:\Theta(-z)-\omega[\rho_{\text{N}}(\varpi)]\:\Theta(z)\}  \nonumber 
\; ,
  \end{eqnarray}
  where $\omega[\rho]=\Phi_{\text{id}}[\rho]+\Phi_{\text{ed}}[\rho]-\mu\rho$ is the grand potential density. For the sharp interface from Eq.~(\ref{dens_sk}) it is sufficient to evaluate
\begin{eqnarray}
\label{eq_gammask}
\gamma_{\text{IN}}[\rho\!\:(z,\varpi)] & = &  \int\mathrm{d}z\;\left\{\Phi_{\text{ed}}(\{n_{\nu}(z)\})  \right. \\ 
  & & \left. -\Phi_{\text{ed}}(\{n_{\nu,\text{I}}\})\:\Theta(-z)-\Phi_{\text{ed}}(\{n_{\nu,\text{N}}\})\:\Theta(z)\right\} \nonumber 
  \end{eqnarray}
as the ideal gas free energy and the density are local quantities. In second order approximation \cite{mcmullen88,mcmullen90} the free energy density can be written as 
\begin{eqnarray}
\label{eq_secondorder} 
\Phi_{\text{ed}}^{(2)}(z) & = & -\frac{1}{2}\int\mathrm{d}\bvec{r}\:\!'\int \mathrm{d}\varpi\int \mathrm{d}\varpi\:\!'\;c^{(2)}\left(\bvec{r}-\bvec{r}\:\!',\varpi,\varpi\:\!'\right)   \cr
 & & \times (\rho\!\:(z,\varpi)-\rho_{\text{I}})\left(\rho\!\:(z\:\!',\varpi\:\!')-\rho_{\text{I}}\right)
  \end{eqnarray}
with the direct correlation function 
 \begin{eqnarray}
 c_{\text{I}}^{(2)}\left(\bvec{r}-\bvec{r}\:\!',\varpi,\varpi\:\!'\right) = -\sum_{\nu, \mu}&&\left.\frac{\partial^2\Phi_{\text{ed}}}{\partial n_{\nu}\partial n_{\mu}}\right|_{\rho=\rho_{\text{I}}}\;  \\  
&& \ \omega^{(\nu)}(\bvec{r}, \varpi)\otimes\omega^{(\mu)}\left(\bvec{r}\:\!', \varpi\:\!'\right) \nonumber 
  \end{eqnarray}
  evaluated at the isotropic coexisting density $\rho_{\text{I}}$. Inserting $\int\mathrm{d}z\;\Phi_{\text{ed}}^{(2)}(z)$ from Eq.~(\ref{eq_secondorder}) and just $\Phi_{\text{ed}}^{(2)}$ for the bulk densities $\rho_{\text{I}}$ and $\rho_{\text{N}}(\varpi)$ into Eq.~(\ref{eq_gammask}) 
leads to 
\begin{equation}
\gamma_{\text{IN}} \simeq \frac{1}{2}\sum_{\nu, \mu}\sigma_{\nu}\left.\frac{\partial^2\Phi_{\text{ed}}}{\partial n_{\nu}\partial n_{\mu}}\right|_{\rho=\rho_{\text{I}}} 
\int\mathrm{d}z \;\tilde{n}_\nu(z) \: \tilde{n}_\mu(-z)
\label{eq_gammaex}
\end{equation}
with the combined weighted densities 
\begin{equation}
 \tilde{n}_\nu(z)=\left(\rho_{\text{N}}\:\omega_{\text{th}}^{(\nu)}(z, \alpha_{\text{N}})-\rho_{\text{I}} \:\omega_{\text{th}}^{(\nu)}(z,0)\right)
 \label{eq_combinedwd}
\end{equation}
 from the thresholded weight functions in Eq.~(\ref{thres_wfO}). The sum runs over all scalar, vectorial and tensorial indices where vectors only contribute if referred to by both $\nu$ and $\mu$. The second order approximation is in good agreement with the direct evaluation of Eq.~(\ref{eq_gamma}) as one can see in Fig.~\ref{fig_ITSK} for spherocylinders of the aspect ratio $l=10$. Shown is the interfacial tension $\gamma_{\text{IN}}$ as function of the tilt angle $\varTheta$ between the interface normal and the nematic director. The minimum $\beta\gamma_{\text{IN}}(L+D)D\approx 0.15$ at $\varTheta\approx \pi/3$, which is obviously an artifact of the sharp-kink profile, was also found in another density functional calculation within the same approximation \cite{holyst88}.  
 Interestingly Fig.~\ref{fig_wofz} reveals that the thresholded weight functions $\omega_{\text{th}}^{(\nu)}(z,\alpha_{\text{N}})$ at a similar tilt angle are nearly identical to those for $\alpha=0$. The resulting uniformly small values of $\tilde{n}_\nu(z)$ in Eq.~(\ref{eq_combinedwd}) could induce the minimum. The calculated interfacial tension is lower than the value obtained in Ref.~\onlinecite{holyst88}.
 Thus it is in reasonable agreement with advanced grand-canonical Monte-Carlo simulations \cite{vink05,wolfsheimer06} and a freely minimized density functional \cite{velasco02}. Introducing a shift $\Delta z$ between the jump of the density and order parameter profile does not change these results significantly as the inset of Fig.~\ref{fig_ITSK} shows. However, it provides the qualitative description of alignment at the isotropic side near the interface. A free minimization of the density functional would certainly lower the values at all angles and would  probably lead to a monotonic decreasing functions as it was found in Refs.~\onlinecite{chen92,SR,velasco02,mcmullen88}. To explore this we will use an evidentially good approximation for the equilibrium profile in Sec.~\ref{sub_ht} but 
emphasize that already such a crude approximation as a sharp-kink interface lead to reasonable values for the isotropic-nematic interfacial tension.

\subsection{Parametrized minimization of a hyperbolic tangent profile} \label{sub_ht}

For a more sophisticated calculation of the interfacial tension we introduce the modulation function
\begin{equation}
h(z)=  \frac{1}{2}\left(1+\tanh \left(\frac{z}{\delta}\right)\right)\, .
\label{eq_modfunct}
\end{equation}
The parameter $\delta$ characterizes the widths of both profiles of the density $\rho\!\:(z)$ and the nematic order parameter $S(z)$ as the system has only one characteristic length scale defined by the correlation length. This can be understood within a Landau-de Gennes expansion \cite{deGennes} done in Sec.~\ref{sec_LdG} for hard particles. The combination of Eqs.~(\ref{eq_profileS}) and (\ref{eq_profileE}) leads to the density profile
\begin{eqnarray}
\rho\!\:(z,\varpi)&=&\left(\rho_{\text{N}}h^2(z)+\rho_{\text{I}}\left(1-h^2(z)\right)\right)\nonumber\\&&g_{\hat{n}}\left(\alpha_{\text{N}}\sqrt{h(z)},\theta, \phi\right)\, .
\label{eq_dens_htL}
 \end{eqnarray} 
\begin{figure} [t]
\centering
\includegraphics[height=0.295\textwidth] {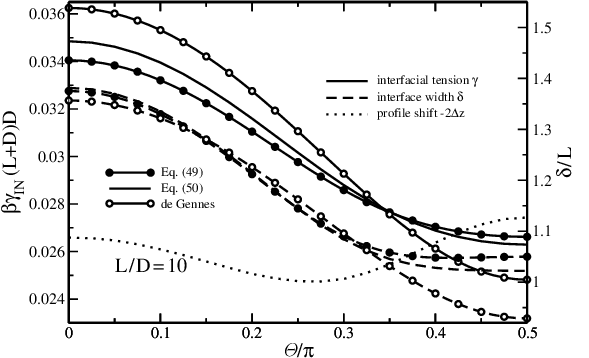}
\caption{Director dependence of the interfacial tension $\gamma_{\text{IN}}$ (left axis, solid lines) as well as width $\delta$ and shift $\Delta z$  (right axis) of the imposed hyperbolic tangent profiles for hard spherocylinders with $l=10$. The interface width $\delta$ (dashed lines) is the only parameter of the profile defined in Eq.~(\ref{eq_dens_htL}) with results indicated by the dots. The dotted line shows the profile shift $\Delta z$ within the two parameter profile from Eq.~(\ref{eq_dens_ht2}). It is renormalized by a factor of $-2$ to fit to this plot. The circles denote the predictions of Landau-de Gennes theory in Sec.~\ref{sec_LdG} which are in good agreement to the DFT results.}
  \label{fig_ITHT05}
\end{figure}
\begin{figure} [b]
\centering
\includegraphics[height=0.295\textwidth]{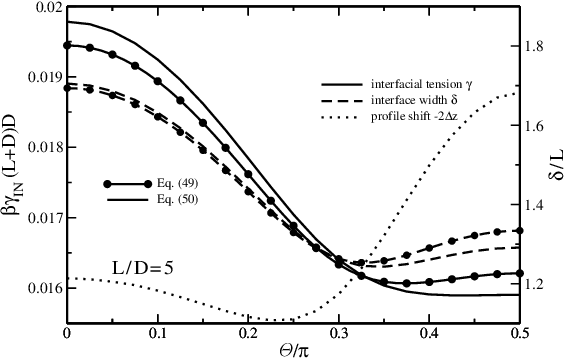}
\includegraphics[height=0.295\textwidth]{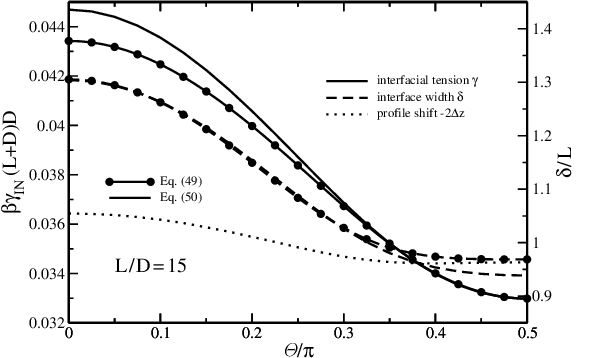}
\includegraphics[height=0.3\textwidth]{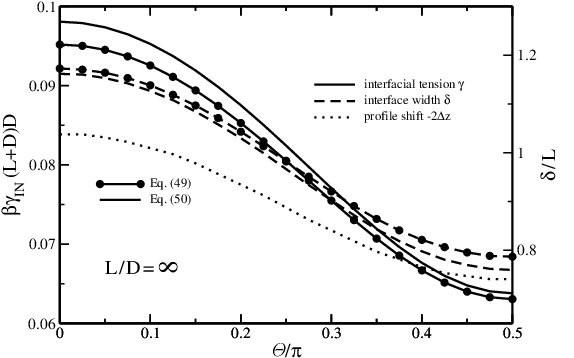}
\caption{ Interfacial tension $\gamma_{\text{IN}}$ and  profile parameters
for (a)  $l=5$, (b) $l=15$ and (c) in the Onsager limit $l=\infty$. Axes and symbols as in Fig.~\ref{fig_ITHT05}. Note that
$\lim_{l\rightarrow\infty}\beta\gamma_{\text{IN}}(L+D)D=\lim_{l\rightarrow\infty}\beta\gamma_{\text{IN}}LD$
and $\delta=\delta_L\:L$}
  \label{fig_ITHT10}
\end{figure}
 Recall from Sec.~\ref{sec_isonemtrans} that the nematic order parameter $S$ is proportional to the squared intrinsic order parameter $\alpha$ of edFMT in first order. Motivated by the usual fit profiles e.g.\ from Refs.~\onlinecite{velasco02,vink05} we also use
\begin{eqnarray}
\rho\!\:(z,\varpi)&=&\left( \rho_{\text{N}}h(z)+\rho_{\text{I}}(1-h(z))\right)\cr && g_{\hat{n}}\left(\alpha_{\text{N}}\sqrt{h(z-\Delta z)},\theta, \phi\right)
\label{eq_dens_ht2}
 \end{eqnarray} 
 as a trial profile. It has the disadvantage of containing an additional parameter $\Delta z$ which denotes the shift between density and order parameter profile. On the other hand the shift obtained in this way can be directly compared to the predictions of simulations. The calculation of the interfacial tension demands the evaluation of the complete expression, Eq.~(\ref{eq_gamma}) in contrast to the sharp-kink approximation. The weighted densities $n_{\nu}(z)$ in $\Phi_{\text{ed}}(\{n_{\nu}(z)\})$ are calculated via Fourier transform of Eq.~(\ref{eq_weighdenorient}) using either Eq.~(\ref{eq_dens_htL}) or Eq.~(\ref{eq_dens_ht2}). In most cases we use a discretization of the $z$-axis with a stepsize of $0.001D$. The number of grid points is adapted to take into account the relevant modulation of the continuous density profile. 
 Minimization is performed with respect to the particular parameters with an accuracy of at least five digits in the interfacial tension. 
 An expansion of $\Phi[\rho]$ to second order as in Eq.~(\ref{eq_gammaex}) has also been done but does not provide any computational bene\-fit.

\begin{figure} [t]
\centering
\includegraphics[height=0.3\textwidth] {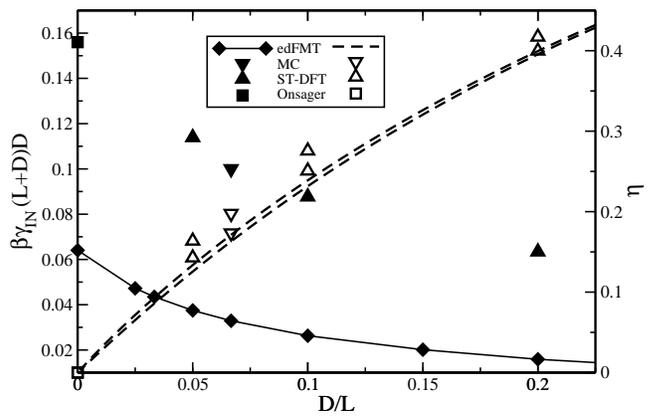}
\caption{Isotropic-nematic interfacial tension $\gamma_{\text{IN}}$ (left axis, filled symbols) and coexisting densities $\eta_{\text{I}}<\eta_{\text{N}}$ (right axis) for different aspect ratios $l$ of hard spherocylinders. The result of this work (lines) is obtained with the interfacial profile given in Eq.~(\ref{eq_dens_ht2}). Comparison is made with grand-canonical Monte-Carlo simulations \cite{vink05,wolfsheimer06,schmid07} (downward triangles), the DFT by Somoza and Tarazona \cite{VMS,velasco02} (upward triangles) and to Onsager theory \cite{SR} (squares). The difference $\delta\eta_{\text{IN}} =\eta_{\text{N}}-\eta_{\text{I}}$ and $\gamma_{\text{IN}}$ from edFMT are significantly smaller.}
  \label{fig_CoexIT}
\end{figure}

 The results with both trial profiles are shown in Figs.~\ref{fig_ITHT05} and \ref{fig_ITHT10} for the aspect ratios of $l=5$, $l=10$, $l=15$ and the Onsager limit. One observes a monotonically decreasing interfacial tension with equilibrium alignment parallel to the interface. The exception of a small increase at high tilt angles for $l=5$ could be an artifact of the parametrized minimization.  
For small tilt angles the trial profile from Eq.~(\ref{eq_dens_htL}) minimizes the interfacial tension, while for higher values including the absolute minimum at $\varTheta=0.5\:\pi$ the two-parameter profile from Eq.~(\ref{eq_dens_ht2}) is a better approximation. At some aspect ratio $40<l<\infty$ the one-parameter profile starts to provide the minimal value for all tilt angles. The difference between those two methods, however, is relatively small. The minimal interfacial tension and the corresponding profile parameters are listed in Table \ref{tab_IN} in addition to the bulk coexistence values. Compared to the sharp-kink profile, the interfacial tension is decreased by one order of magnitude which points out the rigorousness of this approximation. For $l=15$ we obtain a value of $\beta\gamma_{\text{IN}}(L+D)D=0.0329$ which is now significantly smaller than $\beta\gamma_{\text{IN}}(L+D)D=0.10$ from Monte-Carlo simulation\cite{vink05}. 
This difference is to be adressed to a deficiency of the current functional as the errors arising from the simulation are smaller than the symbol size.
The shift $|\Delta z|=0.491\:L$ of the order parameter profile to the isotropic side of the 
interface is in good agreement with the Monte-Carlo value $(0.37\pm0.04)\:L$ and the result $0.45\:L$ from Onsager DFT \cite{SRPhD}. In the Onsager limit we obtain $\beta\gamma_{\text{IN}}(L+D)D=0.0635$ while the most recent numerical study \cite{SR} yields $\beta\gamma_{\text{IN}}(L+D)D=0.156$. The comparison 
in Fig.~\ref{fig_CoexIT} shows the right trend of the interfacial tension $\beta\gamma_{\text{IN}}(L+D)D$ to increase with the aspect ratio $l$. The absolute values, however, are underestimated by a factor between two and four. We further find the normalized interface width $\delta/L$ and the profile shift $|\Delta z|/L$ to be monotonically decreasing functions functions of the aspect ratio $l$.  

\subsection{Discussion}\label{sub_dc}

The pair interaction of two arbitrarily shaped convex hard bodies can be written down exactly as an expansion in tensorial weighted densities, i.e., an infinite series. However, for inhomogeneous systems this is not practicable and leads to the restriction to rank 2 tensors and the introduction of an uncontrolled $\zeta$ parameter \cite{hansengoos09}. The straightforward extrapolation to the excess free energy of dense fluids is based on results  for hard spheres \cite{Rf89}. We point out that it is very difficult to reproduce an appropriate EOS which fulfills the same requirements for hard spherocylinders or arbitrary anisotropic bodies.

The isotropic-nematic interface may be studied analytically by the means of a Landau-de Gennes expansion. The remarkable agreement with the DFT results manifested in Fig.~\ref{fig_ITHT05} suggests a general scaling behavior of the interfacial tension exclusively with different bulk properties according to Eq.~(\ref{ITscale}). This is in agreement with the known weakness of the current density functional to underestimate the difference $\delta\eta_{\text{IN}}$ of the coexisting densities. Figure~\ref{fig_CoexIT} allows a direct comparison of these values. Similar conclusions can be drawn from the nematic order parameter $S_{\text{N}}$. Thus we can use results from the isotropic-nematic transition to predict the surface tension which should be of particular interest for the study of more complicated shapes. 

To study the isotropic-nematic interface we evaluated the present functional in its original form with a fitted value for the $\zeta$ correction \cite{hansengoos09}. 
The results for the interfacial tension suggest a careful examination of this approximation. The change of the $\zeta$ parameter impacts the values of the coexisting densities significantly. However, the small difference $\delta\eta_{\text{IN}}$ as well as the interfacial tension $\gamma_{\text{IN}}$ are both not very sensible to such changes. Considering the phase transition in the Onsager limit we find evidence that it is indeed reasonable to keep the value $\zeta=5/4$ which minimizes the error made for the excluded volume \cite{hansengoos09,hansengoos10} - instead of the fit value $\zeta=1.6$. 

In conclusion, the density functional theory developed in Ref.~\onlinecite{hansengoos09} does not only yield a stable nematic phase but also provides qualitative predictions of the interfacial properties at coexistence. The use of an appropriate continuous trial function for the density profile is completely sufficient to extract all important aspects. The only exception is the explicit shape of the interfacial profiles which may be non-monotonic and show effects of biaxiality as observed in free minimizations \cite{velasco02,SR}. The monotonic director dependence of the interfacial tension \cite{chen92,SR,velasco02,mcmullen88} is reproduced as well as a shift of nematic order to the isotropic side of the interface \cite{chen92,SR,velasco02,vink05,wolfsheimer06}. A free minimization would at most decrease the values of the interfacial tension. 
It is more important to consider the origin of the deviation from the larger simulation values. The third term of the functional is expected to be relevant for the nematic equation of state in addition to the discussed limitations of the $\zeta$ correction. Indeed we have evidence that a different expression will improve the phase behavior. This improvements are quantified in future work where we also need to study higher ordered phases such as smectics to draw general conclusions.

\bigskip 

\noindent 
{\bf Acknowledgment}  

It is a great pleasure to thank Roland Roth for his support and stimulating discussions.  
Thanks to Nelson Rei Bernardino for sharing his expertise about Landau-de Gennes theory and the close collaboration. We also thank Matthieu Marechal for helpful suggestions. Financial support by the DFG under grant Me1361/12 as part of the Research Unit 'Geometry and Physics of Spatial Random Systems' is gratefully acknowledged.

\appendix
\section{Density modulations in one dimension}
\label{appendixA}

\begin{figure} [t]
\centering
\includegraphics[height=0.3\textwidth] {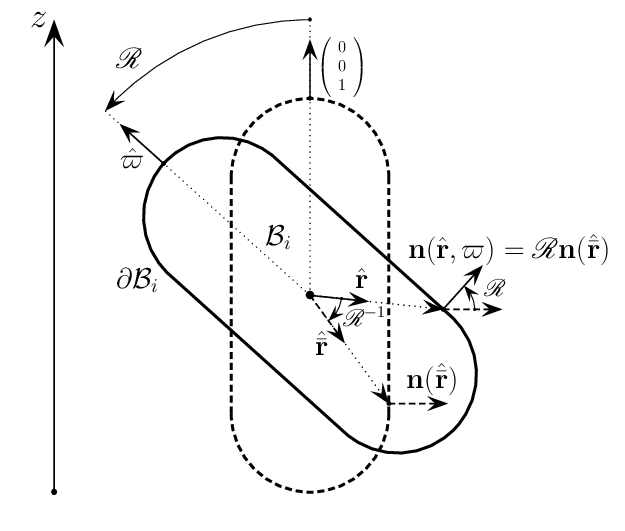}
\caption{Illustration of the orientation-dependence of the weight functions $\omega^{(\nu)}(\bvec{r},\varpi)$ for a convex body $\mathcal{B}_i$. An arbitrarily oriented body (solid line) can be seen as the dashed body oriented in $z$-direction rotated with the matrix $\mathscr{R}(\varpi)$. For any orientation $\varpi$ the unit vector $\hat{\bvec{r}}$ points to a distinct point on the surface $\partial\mathcal{B}_i$.   
According to Eqs.~(\ref{eq_transscal}) and (\ref{eq_transvec}) the surface geometry of the rotated body can be expressed in terms of the geometry of the aligned body and rotation matrices as shown for the normal vector $\bvec{n}(\hat{\bvec{r}},\varpi)$. This suggests a parametrization in body-fixed coordinates $\hat{\bar{\bvec{r}}}:=\mathscr{R}^{-1}(\varpi)\hat{\bvec{r}}$ as illustrated in Fig.~\ref{fig_bodypar}.}
  \label{fig_rotation}
\end{figure}

The general density profile $\rho\!\:(\bvec{r},\varpi) =\rho\!\:(\bvec{r})\:g(\bvec{r},\theta, \phi)$ defines a coordinate system with $\bvec{r}=(x,y,z)^T$. It can be separated into a distribution $\rho(\bvec{r})$ of the centers of mass and an orientational distribution function $g(\bvec{r},\theta, \phi)$ which may have a spatial modulation as well. The orientational average
\begin{equation}
\int \mathrm{d} \varpi=\frac{1}{4\pi}\int_0^{2\pi}\mathrm{d}\phi\int_0^{\pi}\sin\theta\:\mathrm{d}\theta
\label{eq_orav} 
 \end{equation}
is performed with respect to the rotation angles $\theta$ and $\phi$. The orientation matrix
\begin{equation}  
\mathscr{R}(\varpi)=\mat{\cos\phi \cos\theta}{-\sin\phi}{\cos\phi \sin\theta}{\sin\phi \cos\theta}{\cos\phi}{\sin\phi \sin\theta}{-\sin\theta}{0}{\cos\theta}\, 
 \label{rm}
\end{equation}
contains the orientation unit vector  
\begin{equation}
\hat{\varpi}= 
\mathscr{R}(\varpi)\vrx{0}{0}{1}=\vrx{\cos\phi \sin\theta}{\sin\phi \sin\theta}{\cos\theta}\, .
 \end{equation}
The weight functions $\omega^{(\nu)}(\bvec{r},\varpi)$ defined in Eqs.~(\ref{weightfunctions}) and (\ref{weightfunctions2}) can not be parametrized generally as they depend on both position and orientation. In
Fig.~\ref{fig_rotation} we see that this dependence can be decoupled for the scalar
\begin{equation}  
\kappa(\hat{\mathcal{R}})=\kappa\!\left(\mathscr{R}^{-1}(\varpi)\,\hat{\bvec{r}}\right)
\label{eq_transscal}
\end{equation}
and vectorial quantities
\begin{equation}  
\bvec{n}(\hat{\mathcal{R}})=\mathscr{R}(\varpi)\,\bvec{n}\!\left(\mathscr{R}^{-1}(\varpi)\,\hat{\bvec{r}}\right)
\label{eq_transvec}
\end{equation}
which characterize the surface $\partial\mathcal{B}_i$. 
The body-fixed coordinates $\hat{\bar{\bvec{r}}}:=\mathscr{R}^{-1}(\varpi)\,\hat{\bvec{r}}$ allow an explicit parametrization. All vectors present in Eqs.~(\ref{weightfunctions}) and (\ref{weightfunctions2}) need to be transferred according to Eq.~(\ref{eq_transvec}) which gives rise to rotated weight functions $\omega^{(\nu)}_{\mathscr{R}}(\bar{\bvec{r}},\varpi)$.

In the following we consider a cylindrical symmetric density modulation $\rho\!\:(z,\varpi)$. The convolution
\begin{equation}  
\!\int\mathrm{d}\bvec{r}\:\!' \;\omega^{(\nu)}(\bvec{r}-\bvec{r}\:\!',\varpi) \: h(z\:\!') \:=\:\fu{\!\!\int\mathrm{d}\bvec{r}\:\!' \;\omega^{(\nu)}(-\bvec{r}\:\!',\varpi)\:\! h(z\:\!'+z)}{\!\!\int\mathrm{d}z\:\!' \;\omega^{(\nu)}(z-z\:\!',\varpi)\:\! h(z\:\!')}
\label{wallg}
\end{equation}
with an arbitrary function $h(z)$ can be performed in two ways. As illustrated in Fig.~\ref{fig_bodypar} a spherocylinder can be directly parametrized within body-fixed cylindrical coordinates $\bar{\bvec{r}}=(\bar{z}, \bar{\varrho},\bar{\varphi})$ following the substitution $\bvec{r}\:\!'\rightarrow\bvec{r}\:\!'+\bvec{r}$ in the first line of Eq.~(\ref{wallg}). Then we can make the transition 
\begin{equation}
  \omega^{(\nu)}(-\bvec{r}\:\!',\varpi)\rightarrow \omega^{(\nu)}_{\mathscr{R}}(-\bar{\bvec{r}},\varpi)=\sigma_\nu\: \omega^{(\nu)}_{\mathscr{R}}(\bar{\bvec{r}},\varpi)                                                                             
\end{equation}
 to rotated weight functions where the sign function $\sigma_\nu$ is negative only for vectorial weight functions. The rotation 
\begin{equation}
\bvec{r}\:\!'(\bar{\bvec{r}},\varpi)=\mathscr{R}(\varpi)\left( \bar{\varrho}\,\hat{e}_{\bar{\varrho}}+\bar{z}\,\hat{e}_{\bar{z}}\right) 
\label{rot}
  \end{equation} 
of the radial vector $\bar{\bvec{r}}$ results in $\mathrm{d}\bvec{r}\:\!'=\mathrm{d}\bar{\bvec{r}}=\bar{\varrho}\,\mathrm{d}\bar{\varrho}\,\mathrm{d}\bar{\varphi}\,\mathrm{d}\bar{z}$ and $z\:\!'=-\sin\theta\bar{\varrho}\cos\bar{\varphi}+\cos\theta \bar{z}$. That means the orientation dependence is partially transferred to the modulation $h\left(z\:\!'+z\right)$. 
If the five-dimensional integral over $\mathrm{d}\bar{\bvec{r}}$ and $\varpi$ can be solved analytically this straightforward method is convenient. However, this is limited to a few special cases like the calculation of the homogeneous weighted densities of a spherocylinder from Eq.~(\ref{eq_gewdich}). For most inhomogeneous profiles as the sharp-kink in Eq.~(\ref{thres_wf}) this is not possible. Instead of solving those integrals numerically the other conversion in Eq.~(\ref{wallg}) can be applied. It makes use of one-dimensional weight functions $\omega^{(\nu)}(z,\varpi)$ which are calculated in appendix \ref{app_wofz} for a spherocylinder. This reduces the dimension of the integral so that the one-dimensional convolution 
 \begin{equation}  
 \int\mathrm{d}z\:\!'\; \omega^{(\nu)}(z-z\:\!',\varpi) h(z\:\!')=\mathcal{F}T^{-1}\left(\mathcal{F}T\left(\omega^{(\nu)}\right)\ast\mathcal{F}T(h)\right)
\label{eq_convFourier}
\end{equation}
 may be evaluated with a simple multiplication of the Fourier transforms $\mathcal{F}T$ \cite{RevRoland}. The lengthy calculation of $\omega^{(\nu)}(z,\varpi)$ has to be repeated for each different body shape. A similar method can be applied for spherical symmetric geometries. Higher dimensional density modulations need to be handled by a generalized substitution to inner coordinates according to Eq.~(\ref{rot}).

Another important aspect is related to the orientational distribution function $g(\alpha(z),\theta, \phi)$. If the order parameter $\alpha(z)$ is not constant the orientational distribution function is part of the integrand $h(z)$ in Eq.~(\ref{wallg}). This can be implemented straightforwardly in the context of numerical treatment. The density modulation further marks a distinct direction in space. Hence the orientation of the nematic director in the outer coordinate system is no longer arbitrary. The distribution $g(\alpha,\cos\theta)$ from Eq.~(\ref{eq_ordis}) used for homogeneous systems has a maximum at $\theta=0$ which corresponds to a director $\hat{n}=(\sin\varTheta\cos\varPhi,\sin\varTheta\sin\varPhi,\cos\varTheta)^T$ pointing in $z$-direction, i.e.\ $\varTheta=0$. An arbitrary director orientation is equivalent the maximum $g_{\hat{n}}(\alpha,\varTheta, \varPhi)$ of the generalized orientational distribution function
\begin{equation}
g_{\hat{n}}(\alpha,\theta, \phi)=\frac{\alpha}{\mathcal{D}(\alpha)}\exp\left(-\alpha^2\left(1-\cos^2\vartheta\right)\right) \, .
  \end{equation} 
The substituted argument
 \begin{equation}
\cos\vartheta=\sin\varTheta\cos\left(\phi-\varPhi\right)\sin\theta+\cos\varTheta\cos\theta
  \end{equation} 
is the third coordinate of the rotated orientation vector $\hat{\varpi}\rightarrow \mathscr{R}^{-1}(\varTheta,\varPhi)\, \hat{\varpi}$ with respect to the inverse rotation matrix from Eq.~(\ref{rm}) evaluated for the tilt angles $\varTheta$ and $\varPhi$.  Without loss of generality we choose $\varPhi=0$ for a cylindrical symmetric density. For symmetry reasons one always obtains $(\vec{n}_i)_{2}=(\overleftrightarrow{n}_i)_{12}=(\overleftrightarrow{n}_i)_{23}=0$ for the weighted densities in this case. The even more general case of a spatially dependent director orientation addresses to the Frank elastic energy \cite{Frank} which is a topic of future work \cite{PREPEC}.
\begin{figure} [tb]
\centering
\includegraphics[width=0.48\textwidth] {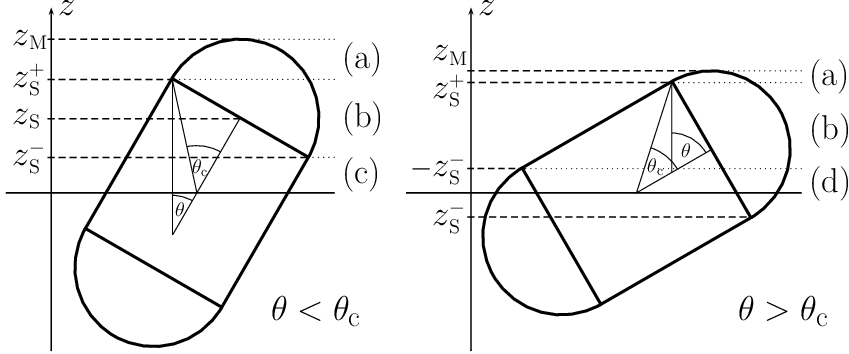} \vspace*{0.1cm}
\caption{Illustration of the four regions on the $z$-axis with different types of the intersection lines between a spherocylinder and parallels to the $xy$-plane. In region $(\text{a})$ there is only one circular intersection of a hemisphere. The cylindrical body sets in to the intersection in region $(\text{b})$. There is no case $(\text{c})$ where only the cylindrical part intersects with an elliptical line for $\theta>\theta_{\text{c}}$ and otherwise no case $(\text{d})$ where all parts contribute. The expressions for the restricting values of $z$ are given in the text.}
\label{fig_regionshsc}
  \end{figure} 

\section{One-dimensional weight functions for spherocylinders}
\label{app_wofz}
The weight functions $\omega^{(\nu)}(z,\varpi)$ of a spherocylinder in a planar geometry are calculated from the intersection lines of the spherocylinder surface with a plane perpendicular to the $z$-axis. We find $\omega^{(\nu)}(z,\varpi)=\sigma_\nu\:\omega^{(\nu)}(-z,\varpi)$ and $\omega^{(\nu)}(z,\varpi)=\omega^{(\nu)}(z,-\varpi)$ from the symmetry of a spherocylinder. Thus only the cases $z>0$ and $0\leq\theta\leq\pi/2$ need to be considered. For convenience we will omit the arguments 
of most functions. From the drawing in Fig.~\ref{fig_regionshsc} one recognizes four different regions with the characteristic functions
\begin{equation}
\fuxyC{\chi_{\text{a}}=1}{\chi_{\text{b}}=1}{\chi_{\text{c}}=1}{\chi_{\text{d}}=1}
\fuxyC{\mbox{if}}{\mbox{if}}{\mbox{if}}{\mbox{if}}
\fuxyC{z_{\text{S}}^+<z<z_{\text{M}}}{\left|z_{\text{S}}^-\right|<z<z_{\text{S}}^+}{0<z<\left|z_{\text{S}}^-\right|\ \wedge \ z_{\text{S}}^->0}{0<z<\left|z_{\text{S}}^-\right|\ \wedge\  z_{\text{S}}^-<0} 
  \end{equation} 
  which are zero otherwise. The condition $z_{\text{S}}^->0$ is equivalent to $\theta<\theta_{\text{c}}=\arctan l$. The boundaries are determined by $z_{\text{M}}=z_{\text{S}}+R$ and $z_{\text{S}}^\pm=z_{\text{S}}\pm R\sin\theta$ with the center $z_{\text{S}}=L\cos\theta/2$ of the upper hemisphere of radius $R=D/2$. With the index $\varsigma\in\{\circ,-,+\}$ and the definitions $Z_{\mp}(z):=z\mp z_{\text{S}}$ and $Z_{\circ}:=Z_{-}$ the partial weight functions of the cylindrical and hemispherical contributions in all regions can be collected separately. For the capping hemispheres one obtains
\begin{eqnarray}\omega^{(\nu)}_{\mathcal{H}}(z,\varpi)&=&\omega^{(\nu)}_{\mathcal{H}^{\circ}}(z) \:\chi_{\text{a}}(z)+\omega^{(\nu)}_{\mathcal{H}^{-}}(z) \:\chi_{\text{b}}(z) \nonumber\\&& +\left(\omega^{(\nu)}_{\mathcal{H}^{-}}(z)+\omega^{(\nu)}_{\mathcal{H}^{+}}(z)\right) \:\chi_{\text{d}}(z)  \end{eqnarray} 
from integrals over the intersecting circles or arcs. The specific contributions read
\begin{eqnarray}
\omega^{(3)}_{\mathcal{H}^{\varsigma}}(z)&=&\frac{R^2-Z_{\varsigma}^2}{2}\left(A_{\mathcal{H}^{\varsigma}}-C_{\mathcal{H}^{\varsigma}}\right)\\
\omega^{(2)}_{\mathcal{H}^{\varsigma}}(z)&=&RA_{\mathcal{H}^{\varsigma}}\\
\vec{\omega}^{(2)}_{\mathcal{H}^{\varsigma}}(z)&=&\vrx{\sqrt{R^2-Z_{\varsigma}^2}\cos\phi\: B_{\mathcal{H}^{\varsigma}}}{ \sqrt{R^2-Z_{\varsigma}^2}\sin\phi\: B_{\mathcal{H}^{\varsigma}}}{Z_{\varsigma}\: A_{\mathcal{H}^{\varsigma}}}
\end{eqnarray}
and
\begin{eqnarray}
\left(\overleftrightarrow{\omega}^{(2)}_{\mathcal{H}^{\varsigma}}(z)\right)_{11}&=& \frac{R^2-Z_{\varsigma}^2}{2R}\left(A_{\mathcal{H}^{\varsigma}}
+\left(2\cos^2\phi-1\right)C_{\mathcal{H}^{\varsigma}}\right) \cr
\left(\overleftrightarrow{\omega}^{(2)}_{\mathcal{H}^{\varsigma}}(z)\right)_{22}&=& \frac{R^2-Z_{\varsigma}^2}{2R}\left(A_{\mathcal{H}^{\varsigma}}
-\left(2\cos^2\phi-1\right)C_{\mathcal{H}^{\varsigma}}\right) \cr 
\left(\overleftrightarrow{\omega}^{(2)}_{\mathcal{H}^{\varsigma}}(z)\right)_{33}&=& \frac{Z_{\varsigma}^2}{R}\:A_{\mathcal{H}^{\varsigma}} \cr
\left(\overleftrightarrow{\omega}^{(2)}_{\mathcal{H}^{\varsigma}}(z)\right)_{12}&=& \frac{R^2-Z_{\varsigma}^2}{R}\cos\phi\sin\phi \:C_{\mathcal{H}^{\varsigma}} \cr 
\left(\overleftrightarrow{\omega}^{(2)}_{\mathcal{H}^{\varsigma}}(z)\right)_{13}&=& \frac{Z_{\varsigma}\sqrt{R^2-Z_{\varsigma}^2}}{R}\cos\phi\: B_{\mathcal{H}^{\varsigma}} \cr 
\left(\overleftrightarrow{\omega}^{(2)}_{\mathcal{H}^{\varsigma}}(z)\right)_{23}&=& \frac{Z_{\varsigma}\sqrt{R^2-Z_{\varsigma}^2}}{R}\sin\phi\: B_{\mathcal{H}^{\varsigma}} 
\end{eqnarray}
with the short notations
\begin{eqnarray}
A_{\mathcal{H}^{\circ}}&=&2\pi\ ,\ \  A_{\mathcal{H}^{\mp}}=2\arccos\left(\mp Z_{\mathcal{H}^{\mp}}\right) \, , \cr
B_{\mathcal{H}^{\circ}} & =&0 \ \ \;,\ \  B_{\mathcal{H}^{\mp}} =\pm 2\sqrt{1-Z_{\mathcal{H}^{\mp}}^2} \, ,\\
C_{\mathcal{H}^{\circ}}& =& 0\ \ \;,\ \ C_{\mathcal{H}^{\mp}}=\mp 2Z_{\mathcal{H}^{\mp}}\sqrt{1-Z_{\mathcal{H}^{\mp}}^2} \nonumber
\end{eqnarray}
and
\begin{equation}
Z_{\mathcal{H}^{\mp}}=\frac{Z_{\mp}}{\sqrt{R^2-Z_{\mp}^2}\tan\theta}\, .
\end{equation}
The partial weight functions from the elliptical segments of the cylindrical parts read
\begin{eqnarray}\omega^{(\nu)}_\mathcal{C}(z,\varpi)&=& \omega^{(\nu)}_{\mathcal{C}^{-}}(z) \:\chi_{\text{b}}(z)+\omega^{(\nu)}_{\mathcal{C}^{\circ}} \:\chi_{\text{c}}(z) \nonumber\\ && +\left(\omega^{(\nu)}_{\mathcal{C}^{-}}(z)-\omega^{(\nu)}_{\mathcal{C}^{+}}(z)\right)\:\chi_{\text{d}}(z) \,.\end{eqnarray} 
 We find the parameters
\begin{eqnarray}
A_{\mathcal{C}^{\circ}}&=&\frac{2\pi}{\cos\theta}\, ,\ \  A_{\mathcal{C}^{\mp}}=\frac{2}{\cos\theta}\arccos\left(Z_{\mathcal{C}^{\mp}}\right) \, , \cr 
B_{\mathcal{C}^{\circ}} & =&0 \ \ \ \ \ \, ,\ \  B_{\mathcal{C}^{\mp}} =-2\sqrt{1-Z_{\mathcal{C}^{\mp}}^2} \, ,\\
C_{\mathcal{C}^{\circ}}& =& 0\ \ \ \ \ \, ,\ \ C_{\mathcal{C}^{\mp}}=\frac{2}{\cos\theta}Z_{\mathcal{C}^{\mp}}\sqrt{1-Z_{\mathcal{C}^{\mp}}^2} \nonumber
\end{eqnarray}
with
\begin{equation} 
Z_{\mathcal{C}^{\mp}}=\frac{Z_{\mp}}{R\sin\theta}\, .
\end{equation}
and obtain
\begin{eqnarray}
\omega^{(3)}_{\mathcal{C}^{\varsigma}}(z)&=&\frac{R^2}{2}\left(A_{\mathcal{C}^{\varsigma}}-C_{\mathcal{C}^{\varsigma}}\right)\\
\omega^{(2)}_{\mathcal{C}^{\varsigma}}(z)&=&R\:A_{\mathcal{C}^{\varsigma}}\\
\vec{\omega}^{(2)}_{\mathcal{C}^{\varsigma}}(z)&=&R\vrx{\cos\phi}{\sin\phi}{-\tan\theta}B_{\mathcal{C}^{\varsigma}}\, ,
\end{eqnarray}
\begin{eqnarray}
\left(\overleftrightarrow{\omega}^{(2)}_{\mathcal{C}^{\varsigma}}(z)\right)_{11}&=& \frac{R}{2} \left( \left(1-\sin^2\theta\cos^2\phi \right) A_{\mathcal{C}^{\varsigma}} \right.\cr&&\left.\ \ \ \ \, +\left( \cos^2\theta\cos^2\phi-\sin^2\phi \right) C_{\mathcal{C}^{\varsigma}}\right)  \cr
\left(\overleftrightarrow{\omega}^{(2)}_{\mathcal{C}^{\varsigma}}(z)\right)_{22}&=& \frac{R}{2} \left( \left(1-\sin^2\theta\sin^2\phi \right) A_{\mathcal{C}^{\varsigma}} \right.\cr&&\left.\ \ \ \ \, +\left( \cos^2\theta\sin^2\phi-\cos^2\phi \right) C_{\mathcal{C}^{\varsigma}}\right)  \cr
\left(\overleftrightarrow{\omega}^{(2)}_{\mathcal{C}^{\varsigma}}(z)\right)_{33}&=& \frac{R}{2} \left( \sin^2\theta \: A_{\mathcal{C}^{\varsigma}}+ \sin^2\theta \:C_{\mathcal{C}^{\varsigma}}\right)  \cr
\left(\overleftrightarrow{\omega}^{(2)}_{\mathcal{C}^{\varsigma}}(z)\right)_{12}&=& \frac{R}{2} \left( -\sin^2\theta\sin\phi\cos\phi \: A_{\mathcal{C}^{\varsigma}} \right.\cr&&\left.\ \ \ \ \, + (1+\cos^2\theta)\sin\phi\cos\phi\: C_{\mathcal{C}^{\varsigma}}\right)  \cr
\left(\overleftrightarrow{\omega}^{(2)}_{\mathcal{C}^{\varsigma}}(z)\right)_{13}&=& \frac{R}{2} \left( -\sin\theta\cos\theta\cos\phi \: A_{\mathcal{C}^{\varsigma}} \right.\cr&&\left.\ \ \ \ \, -\sin\theta\cos\theta\cos\phi \: C_{\mathcal{C}^{\varsigma}}\right)  \cr
\left(\overleftrightarrow{\omega}^{(2)}_{\mathcal{C}^{\varsigma}}(z)\right)_{23}&=& \frac{R}{2} \left( -\sin\theta\cos\theta\sin\phi \: A_{\mathcal{C}^{\varsigma}}  \right.\cr&&\left.\ \ \ \ \, -\sin\theta\cos\theta\sin\phi\: C_{\mathcal{C}^{\varsigma}}\right) 
\end{eqnarray} and
\begin{eqnarray}
\left(\overleftrightarrow{\omega}^{(1)}_{\mathcal{C}^{\varsigma}}(z)\right)_{11}&=&\frac{R}{2}\left(\left( 1-3\sin^2\theta\cos^2\phi \right) A_{\mathcal{C}^{\varsigma}}\right.\cr&&\left.\ \ \ \ \, -\left(\cos^2\theta\cos^2\phi-\sin^2\phi\right) C_{\mathcal{C}^{\varsigma}}\right) \cr
\left(\overleftrightarrow{\omega}^{(1)}_{\mathcal{C}^{\varsigma}}(z)\right)_{22}&=&\frac{R}{2}\left(\left( 1-3\sin^2\theta\sin^2\phi \right) A_{\mathcal{C}^{\varsigma}}\right.\cr&&\left.\ \ \ \ \, -\left(\cos^2\theta\sin^2\phi-\cos^2\phi\right) C_{\mathcal{C}^{\varsigma}}\right) \cr
\left(\overleftrightarrow{\omega}^{(1)}_{\mathcal{C}^{\varsigma}}(z)\right)_{33}&=&\frac{R}{2}\left(\left( 1-3\cos^2\theta \right) A_{\mathcal{C}^{\varsigma}}-\sin^2\theta \:C_{\mathcal{C}^{\varsigma}}\right) \cr
\left(\overleftrightarrow{\omega}^{(1)}_{\mathcal{C}^{\varsigma}}(z)\right)_{12}&=&\frac{R}{2}\left( -3\sin^2\theta\sin\phi\cos\phi \: A_{\mathcal{C}^{\varsigma}}\right.\cr&&\left.\ \ \ \ \, -(1+\cos^2\theta)\sin\phi\cos\phi \:C_{\mathcal{C}^{\varsigma}}\right) \cr
\left(\overleftrightarrow{\omega}^{(1)}_{\mathcal{C}^{\varsigma}}(z)\right)_{13}&=&\frac{R}{2}\left( -3\sin\theta\cos\theta\cos\phi \: A_{\mathcal{C}^{\varsigma}}\right.\cr&&\left.\ \ \ \ \, +\sin\theta\cos\theta\cos\phi \:C_{\mathcal{C}^{\varsigma}}\right) \cr
\left(\overleftrightarrow{\omega}^{(1)}_{\mathcal{C}^{\varsigma}}(z)\right)_{23}&=&\frac{R}{2}\left( -3\sin\theta\cos\theta\sin\phi \: A_{\mathcal{C}^{\varsigma}}\right.\cr&&\left.\ \ \ \ \, +\sin\theta\cos\theta\sin\phi \:C_{\mathcal{C}^{\varsigma}}\right) \, .
\end{eqnarray}
 The complete expressions for all weight functions are 
  \begin{eqnarray}
  \omega^{(0)}(z,\varpi)&=&\frac{1}{\pi D^2}\:\omega^{(2)}_{\mathcal{H}}(z,\varpi)\cr
  \omega^{(1)}(z,\varpi)&=&\frac{1}{2\pi D}\:\omega^{(2)}_{\mathcal{H}}(z,\varpi)+\frac{1}{4\pi D}\:\omega^{(2)}_\mathcal{C}(z,\varpi) \cr
  \vec{\omega}^{(1)}(z,\varpi)&=&\frac{1}{2\pi D}\:\vec{\omega}^{(2)}_{\mathcal{H}}(z,\varpi)+\frac{1}{4\pi D}\:\vec{\omega}^{(2)}_\mathcal{C}(z,\varpi) \cr
  \overleftrightarrow{\omega}^{(1)}(z,\varpi)&=&\frac{1}{4\pi D}\:\overleftrightarrow{\omega}^{(1)}_\mathcal{C}(z,\varpi)\cr
  \omega^{(\nu)}(z,\varpi)&=&\omega^{(\nu)}_{\mathcal{H}}(z,\varpi)+\:\omega^{(\nu)}_\mathcal{C}(z,\varpi) 
\label{eq_wfzc}
  \end{eqnarray}
   with the latter equation for $\omega^{(3)}$, $\omega^{(2)}$, $\vec{\omega}^{(2)}$ and $\overleftrightarrow{\omega}^{(2)}$. Note that, for a fixed orientation $\varpi$, the thresholded weight functions $\omega_{\text{th}}^{(\nu)}(z, \varpi)=\int_{-\infty}^{z}\mathrm{d}z\:\!'\; \omega^{(\nu)}(z\:\!',\varpi)$ from Eq.~(\ref{thres_wf}) are the integral functions of these one-dimensional weight functions. The weighted densities $n_\nu(z)$ are either evaluated directly by Fourier transforms as in Eq.~(\ref{eq_convFourier}) or can be further simplified to 
  \begin{eqnarray} 
 n_\nu(z)= 
\frac{1}{2\pi}&&\int_0^{2\pi}\mathrm{d}\phi\int_0^{\frac{\pi}{2}}\sin\theta\:\mathrm{d}\theta\int_0^\infty\mathrm{d}z\:\!' \;\omega^{(\nu)}(z\:\!',\varpi)
\: \cr \times &&\left(\sigma_\nu\:\rho\!\:(z\:\!'+z,\varpi) +\rho\!\:(-z\:\!'+z,\varpi)\right)
 \label{eq_wdz}
\end{eqnarray}
using the symmetry of a spherocylinder. For infinitely long rods the width $\delta=\delta_LL$ of the density modulation defined in Eq.~(\ref{eq_modfunct}) becomes infinitely wide. The substitution $z\rightarrow z_LL$ allows a scaling of Eq.~(\ref{eq_wdz}). The dimensionless concentration $c$ defined by
\begin{equation} 
 \rho\!\:(z)=\rho\!\:(z_LL)=b\!\:c\!\:(z_L)=\frac{4}{L^2D\pi}\!\:c\!\:(z_L)
\end{equation}
remains finite in the Onsager limit. The Onsager excess free energy
\begin{equation}
\label{eq_PhiedOns}
   \lim\limits_{\substack{\rho\rightarrow 0 \\l\rightarrow\infty} }   
   l^2 \Phi_{\text{ed}}\left(\{n_{\nu}\}\right) = \lim\limits_{\substack{\rho\rightarrow 0 \\l\rightarrow\infty} }   l^2\left(n_1 n_2 - \zeta \mathrm{Tr}[\overleftrightarrow{n}_1 \overleftrightarrow{n}_2]\right)
\end{equation} 
is constituted of four scalar or tensorial weighted densities
\begin{eqnarray} 
 \lim_{l\rightarrow\infty}\frac{n_\nu(z)}{lb}=D&&\int\mathrm{d}\varpi\;\omega_{\mathcal{C}^{\circ}}^{(\nu)}(\varpi)\int_0^{\frac{\cos\theta}{2}}\mathrm{d}z_L\:\!' \\\nonumber \times && \left(c\!\:(z_L\:\!'+z_L,\varpi) +c\!\:(-z_L\:\!'+z_L,\varpi)\right)\, .
\end{eqnarray} 
where only the cylindrical parts of region $(c)$ scale with $l$. 
Note that the term $n_0n_3$ does not appear in Eq.~(\ref{eq_PhiedOns}) as the weight function $\omega^{(0)}$ is only non-zero for the capping hemispheres.


\begin{thebibliography}{99}
\bibitem{RemDim} Note here that the dimension of $n_{\nu}$ is (length)$^{\nu-3}$ and the dimension of $\Phi$ is (length)$^{-3}$.
\end{thebibliography}
\end{document}